\documentclass[journal,a4paper,onecolumn, draftcls, 12pt]{IEEEtran}

\IEEEoverridecommandlockouts

\newcommand\hmmax{0}
\newcommand\bmmax{0}

\usepackage{graphicx}
\usepackage{Files/LVcolours} 
\usepackage{tikz}
\usepackage{pgfplots}

\usepackage{amsmath}
\usepackage{amsfonts, amssymb}
\usepackage{mathtools}
\usepackage{cite}
\usepackage[nolist]{acronym}
\usepackage{amsmath}

\usepackage{booktabs} 		
\usepackage{diagbox}		
\usepackage{upgreek}
\usepackage{bbm}
\usepackage{Files/bm}
\usepackage{Files/winsnotation}
\usepackage{Files/Symbols_OPL}

\usepackage[ruled,vlined,noresetcount]{algorithm2e}

\usepackage{setspace}	 	
\usepackage{comment}
\usepackage{physics}
\usepackage{multirow}
\usepackage{ragged2e}
\usepackage{textcomp}

\usepackage{array}
\newcolumntype{C}[1]{>{\centering\arraybackslash}p{#1}}

\makeatletter
\def\endthebibliography{%
  \def\@noitemerr{\@latex@warning{Empty `thebibliography' environment}}%
  \endlist
}
\makeatother
\usepackage{amsthm}
\theoremstyle{definition}

\newcommand{\Syndrome}[0]{\V{s}}

\pgfplotsset{compat=1.17}

\begin{document}

\date{2024}

\title{Reliable Quantum Communications based on Asymmetry in Distillation and Coding}

\author{%
  \IEEEauthorblockN{Lorenzo Valentini, René Bødker Christensen, Petar Popovski, Marco Chiani}
\thanks{Lorenzo Valentini and Marco Chiani are with CNIT/WiLab, DEI, University of Bologna, Italy. Email: \{lorenzo.valentini13, marco.chiani\}@unibo.it. 
René Bødker Christensen is with the Electronic Systems Department, Aalborg University, Denmark and the Department of Mathematical Sciences, Aalborg University, Denmark. Email: rene@math.aau.dk.
Petar Popovski is with the Electronic Systems Department, Aalborg University, Denmark. Email: petarp@es.aau.dk.
This paper was supported in part by the Villum Investigator Grant “WATER” from the Velux Foundations, Denmark, and by the European Union - Next Generation EU, PNRR project, Italy, PRIN n. 2022JES5S2. 
}
}

\maketitle







\begin{abstract}
The reliable provision of entangled qubits is an essential precondition in a variety of schemes for distributed quantum computing. This is challenged by multiple nuisances, such as errors during the transmission over quantum links, but also due to degradation of the entanglement over time due to decoherence. The latter can be seen as a constraint on the latency of the quantum protocol, which brings the problem of quantum protocol design into the context of latency-reliability constraints. We address the problem through hybrid schemes that combine: (1) indirect transmission based on teleportation and distillation; (2) direct transmission, based on \ac{QEC}. The intuition is that, at present, the quantum hardware offers low fidelity, which demands distillation; on the other hand, low latency can be obtained by \ac{QEC} techniques. It is shown that, in the proposed framework, the distillation protocol gives rise to asymmetries that can be exploited by asymmetric \ac{QECC}, which sets the basis for unique hybrid distillation and coding design. 
Our results show that ad-hoc asymmetric codes give, compared to conventional \ac{QEC}, a performance boost and codeword size reduction both in a single link and in a quantum network scenario.
\end{abstract}

\begin{keywords}
Quantum communication, quantum distillation, entanglement, asymmetric channels, asymmetric quantum error correction.
\end{keywords}



\maketitle 
\markboth{Accepted on IEEE Transactions on Quantum Engineering}{}

\begin{acronym}
\small
\acro{AWGN}{additive white Gaussian noise}
\acro{BCH}{Bose–Chaudhuri–Hocquenghem}
\acro{CDF}{cumulative distribution function}
\acro{CRC}{cyclic redundancy code}
\acro{CSS}{Calderbank-Shor-Steane}
\acro{EPR}{Einstein–Podolsky–Rosen}
\acro{LDPC}{low-density parity-check}
\acro{ML}{maximum likelihood}
\acro{MWPM}{minimum weight perfect matching}
\acro{QECC}{quantum error correcting code}
\acro{QEC}{quantum error correction}
\acro{QKD}{quantum key distribution}
\acro{PDF}{probability density function}
\acro{PMF}{probability mass function}
\acro{MPS}{matrix product state}
\acro{WEP}{weight enumerator polynomial}
\acro{WE}{weight enumerator}
\end{acronym}
\setcounter{page}{1}

\section{Introduction}

The evolution of the quantum Internet happens in symbiosis with classical communication and the existing Internet technology, leading to a number of interesting research challenges~\cite{WehElkHan:18,Cac19:QInternet,Pom22:experimentalQI}.
There are two principal types of applications for quantum communication: \emph{(a)} enhancing already existing services such as \acl{QKD} \cite{BenBra84:QKD} or super-dense coding \cite{BenWie92:SuperDenseCoding}; \emph{(b)} enabling new services, such as distributed quantum computing \cite{Bal22:MultiCoreQComp,SunGup22:QcircuitsOverQNet,Ferrari2023:DistrQC} or remote processing of quantum sensing data \cite{Deg17:ReviewQSensing,Bi19:QRemoteSensing}. 
The applications in \emph{(a)} convey classical information, while the applications in \emph{(b)} convey quantum information. 
A quantum network can be defined as a collection of nodes that are able to exchange qubits and distribute entangled states among the network nodes~\cite{irtf-qirg-principles-11}. Two possible ways to exchange quantum information are the indirect transmission through quantum teleportation protocol \cite{Ben93:teleporting}, and direct transmission of qubits exploiting \ac{QEC} \cite{Fow10:QcommWithSurf}.

For \emph{indirect transmission}, the key building block to achieve reliable communication over teleportation is the distribution of \ac{EPR} entangled pairs or other types of entangled states \cite{GreHorZei89:GHZ,Hon04:GHZAndWStates,Pom21:ExperimentalQNetwork}. 
Since such a distribution is affected by imperfections \cite{Cir97:EntanGen, Chi05:EntanGen, Loo06:EntaGen, Uph16:EntaGen, Hu21:EntaGen, RoadmapIntegratedQPhotonics22}, distillation protocols \cite{Ben96:purification, Deu96:purification} have been developed to increase the fidelity of the shared \ac{EPR} pairs. 
In this way, it is possible to make the communication reliable at the cost of an increase in both latency and usage of qubit resources.

There are three basic schemes for heralded entanglement generation
\cite{CacCalVan:20}: $i)$ \emph{at source}, where the transmitter is in charge of generating and sharing the entanglement;  $ii)$ \emph{at mid-point}, where the entanglement source is in between the transmitter and the receiver; $iii)$ \emph{at both end-points}, where the entanglement sharing is addressed cooperatively by the transmitter and receiver. These schemes have resulted in several proposals for quantum Internet protocol stacks.  
For example, in \cite{Van08:Qrepeater} a protocol stack based on \emph{at source} distribution was proposed, while in \cite{Weh19:QStackCoopHeralding, KozWeh20:QStackCoopHeralding} the authors work with distribution based on \emph{at both end-points}.

\begin{figure}[t]
    \centering
    \includegraphics[width = 0.8\columnwidth]{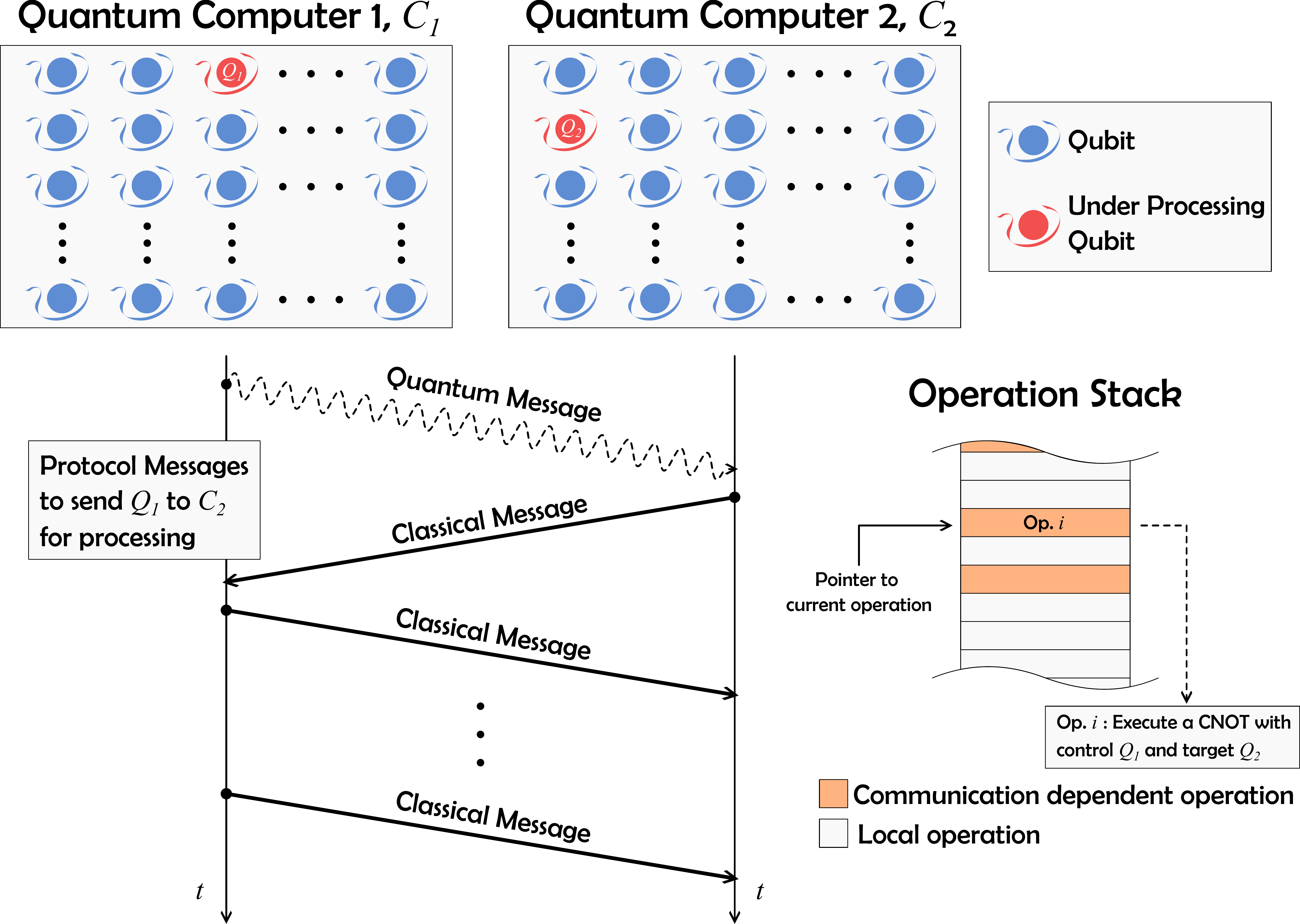}
    \caption{\justifying A motivating example of distributed quantum computation. Two quantum computers $C_1$ and $C_2$ share a common task represented by a sequential operation stack. Following certain scheduling related to the computation, whenever two qubits of different quantum computers require an interaction, a quantum communication link is used to send one of these qubits to the other, execute the operation, and send it back.
    \label{fig:ToyExampleDistributedQuantumComputing}}
\end{figure}

The toy example from  Fig.~\ref{fig:ToyExampleDistributedQuantumComputing} features an indirect communication among two quantum computers using \emph{at source} entanglement distribution.
In particular, the timing diagram of Fig.~\ref{fig:ToyExampleDistributedQuantumComputing} illustrates the overall latency for sending a qubit $Q_1$ from a quantum computer $C_1$ to another computer $C_2$. 
First, a quantum message is sent from $C_1$ to $C_2$, containing a number of qubits (halves of \ac{EPR} pairs). 
Due to noise, the shared pairs will typically no longer be fully entangled, and therefore not suitable to be used directly for teleportation. 
Thus, a chain of classical messages is exchanged to improve the reliability of the pairs, according to distillation protocols \cite{Ben96:purification}. On the flip side, this exchange of messages will also increase the overall latency.
Finally, a (classical) teleportation message is sent to convey the quantum information. 
Note that, due to decoherence, the fidelity of the shared pairs will degrade over time; this is modeled as a constraint on the protocol latency. 

For the example on Fig.~\ref{fig:ToyExampleDistributedQuantumComputing},
\emph{reliability} is the ability to preserve the transmitted state along the path to the receiver, while \emph{latency} is the time between the intention to transmit a qubit and its actual reception.
Suppose one has to execute a quantum algorithm involving many qubits, such as in the factoring algorithm \cite{Sho94:Factoring}. 
We recall that any quantum computation can be implemented by adopting a finite set of universal quantum gates, operating on one or two qubits.
The possible scheduling of operations is represented by the stack in Fig.~\ref{fig:ToyExampleDistributedQuantumComputing}.
When the quantum memory of a single quantum computer is not sufficient for the algorithm, we could consider exploiting multiple quantum computers, for example, two of them indicated as $C_1$ and $C_2$ in Fig.~\ref{fig:ToyExampleDistributedQuantumComputing}. 
In order to perform double qubit operators in a distributed manner, the two quantum computers should be able to reliably exchange qubits.
Assume that, at a given time, a qubit $Q_1$ of $C_1$ has to interact with a qubit $Q_2$ of $C_2$. 
Thus, $C_1$ sends its qubit to $C_2$ using an indirect communication based on distillation and teleportation as previously discussed (see also the timing diagram).
Computer $C_2$ may now execute the local operation on $Q_1$ and $Q_2$.
We note that a higher reliability constraint demands a larger number  of distillation (classical) messages, resulting in increased latency of the communication protocol and directly affecting the algorithm's computational time. Note that reliability constraint is crucial since communication errors lead to reset/restart of the algorithm from the beginning. 
From this simple example, we find a motivation to search for strategies reaching the same reliability as the distillation protocol, while minimizing the number of exchanged classical messages.

Regarding \emph{direct} communication, in \cite{Fow10:QcommWithSurf} the authors consider a quantum link in which the information qubits are conveyed by surface codes \cite{BraKit98:Surface,ValFor23:Surface, For24:XZZXRotAnalysis}. When \ac{QEC} is present, it is also possible to piggyback classical information over quantum information \cite{ChiConWin:20}. In general, we can classify \acp{QECC} into symmetric \cite{Sho:95, Laf:96, Gra07:Codes} and asymmetric \cite{Sar:2009, ChiVal:20a}. 
A symmetric $[[n,k,d]]$ \ac{QECC} encodes $k$ information qubits into $n$ physical qubits and is able to correct up to $t = \lfloor(d-1)/2 \rfloor$ generic errors (i.e., Pauli $\M{X}$, $\M{Z}$, or $\M{Y}$) on the coded qubits. 
On the other hand, an asymmetric $[[n,k]]$ \ac{QECC} with $(e_\mathrm{g}, e_\mathrm{Z})$ error correction capability \cite{ChiVal:20a} is able to correct up to $e_\mathrm{g}$ generic errors and $e_\mathrm{Z}$ particular errors, in this case Pauli $\M{Z}$ ones. 
Other asymmetric codes can be defined as $[[n,k,d_\mathrm{X}/d_\mathrm{Z}]]$ due to the fact that they are able to correct $t_\mathrm{X} = \lfloor(d_\mathrm{X}-1)/2 \rfloor$ Pauli $\M{X}$ errors and $t_\mathrm{Z} = \lfloor(d_\mathrm{Z}-1)/2 \rfloor$ Pauli $\M{Z}$ errors (e.g., \acl{CSS} codes \cite{Ste96:CSS, CalSho96:CSS, Ste99:CSS, Sar:2009}).
In quantum communication, given a fixed error probability on the transmitted qubits, \ac{QEC}-based communication schemes provide higher reliability than uncoded schemes, whenever the initial error probability is below a certain code-dependent threshold.
In general, this initial error probability is termed \emph{fidelity} and is technology-dependent.
Furthermore, in direct communication protocols reliability comes at the cost of quantum memory size through~coding.

This idea to subdivide quantum communication protocols into direct and indirect ones also appears in \cite{Mur16:QnetworkGenerations}, where networks are categorized into ``generations'' based on the employed error management scheme. 
Some of them deal with errors by the mean of distillation, others using \ac{QEC}. 
In this context, we locate our paper in between these two kinds of generations and investigate hybrid schemes adopting both distillation and \ac{QEC} techniques. 
In particular, we assume that local errors (i.e., quantum gate errors and quantum measurements) are negligible compared to non-local errors (i.e., fidelity reduction due to transmission). 
In such a scenario, direct communication is not yet feasible, but at each node of the network \ac{QEC} can be exploited.
The justification of using \ac{QEC} schemes (both hybrid or purely \ac{QEC}-based) is the latency reduction that can be obtained compared to uncoded ones. 
In fact, to achieve the same target reliability, the uncoded scheme requires more distillation steps, expressed as transmissions of classical messages. 
On the other hand, it is also possible to fix the latency by presetting the number of distillation steps and obtain a gain in reliability using \acp{QECC}.
Furthermore, in this new framework, we point out that the distillation protocol gives rise to asymmetries which can be exploited by asymmetric \ac{QECC}s \cite{Sar:2009,ChiVal:20a}, and therefore, we can have a joint distillation and coding design.
Besides the discussed example in Fig.~\ref{fig:ToyExampleDistributedQuantumComputing}, applications can be represented by mid-generation error management schemes for quantum internet, adaptive \ac{QEC} for reliable communication over quantum networks, multi-core quantum computing \cite{Bal22:MultiCoreQComp}, construction of quantum graph state \cite{Hei04:MultipartyGraphStates,Hei06:GraphStates, Mei19:DistribGraphStates} over networks, and many others.

The key contributions of the paper can be summarized as follows:
\begin{itemize}
    \item Proposal of a hybrid distillation-\ac{QEC} scheme which exploits the asymmetries arising from distillation protocols;
    \item Evolution analysis of the equivalent quantum channel parameters of a distillation protocol to find out exploitable asymmetries;
    \item Proposal and design of an improved quantum network protocol based on scheduling policies for swapping and distillation.
    \item Demonstration that ad-hoc asymmetric codes can provide, compared to conventional \ac{QEC}, a performance boost and codeword size reduction both in a single link and in a network scenario;
\end{itemize}
This paper is organized as follows. Section~\ref{sec:preliminary} introduces preliminary concepts and models together with some background material. Section~\ref{sec:mainContribution} focuses on the main contributions of the paper. Numerical results are shown in Section~\ref{sec:NumRes}. Finally, conclusions are drawn in Section~\ref{sec:conclusions}.

\emph{Notation}: 
throughout the paper, capital bold letters denote matrices. We adopt the bra-ket notation to indicate vectors representing quantum states. The ket vector $\ket{\psi}$ is a column vector with complex coefficients, while the bra vector $\bra{\psi}$ is its complex conjugate.

\section{Preliminaries and Background}
\label{sec:preliminary}

A qubit is an element of the two-dimensional Hilbert space $\mathcal{H}^{2}$, with orthonormal basis $\ket{0}$ and $\ket{1}$ \cite{Rie00:introduction,NieChu:10}. 
The Pauli operators $\M{I}, \M{X}, \M{Z}$, and $\M{Y}$, are defined by  $\M{I}\ket{a}=\ket{a}$, $\M{X}\ket{a}=\ket{a\oplus 1}$, $\M{Z}\ket{a}=(-1)^a\ket{a}$, and $\M{Y}\ket{a}=i(-1)^a\ket{a\oplus 1}$ for $a \in \lbrace0,1\rbrace$ where $\oplus$ is the XOR operation. These operators either commute or anticommute with each other. 
Other useful single qubit gates which are used in this paper are the Hadamard gate described as $\M{H} = (\M{X} + \M{Z}) / \sqrt{2}$ and the $x$-axis rotation gate $\M{R}_x(\theta) = \cos(\theta/2) \M{I} - i \sin(\theta/2) \M{X}$.
We use the notation $\ket{\Phi^{\pm}} = (\ket{00} \pm \ket{11})/\sqrt{2}$ and $\ket{\Psi^{\pm}} = (\ket{01} \pm \ket{10})/\sqrt{2}$ for two-qubit Bell's states or, equivalently, \ac{EPR} pairs. 
A CNOT gate is a quantum gate that acts on two qubits, one referred to as \emph{control} and the other as \emph{target}. 
In particular, if the control qubit is in state $\ket{1}$, then the target qubit is inverted, otherwise, nothing happens according to $\ket{c}\ket{t} \mapsto \ket{c}\ket{t \oplus c}$.
Considering an $n$-qubit system, we indicate with $\M{Z}_j$ a Pauli $\M{Z}$ operator acting on the $j$-th qubit. 
Similarly for a CNOT gate, with $\M{{CX}}_{j,k}$ we indicate that the operator has the $j$-th qubit as control and the $k$-th qubit as target.
We define a mixed state as a distribution over quantum states, $\{ p_i, \ket{\psi_i} \}$, meaning that with probability $p_i$ the system is in state $\ket{\psi_i}$. We represent the state of the quantum system using the density matrix representation $\M{\rho} = \sum_i p_i \ket{\psi_i} \bra{\psi_i}$.

The quantum teleportation protocol is an algorithm that transfers a quantum state from one qubit to another at the cost of two classical bits and an \ac{EPR} pair \cite{Ben93:teleporting}. 
The procedure can be summarized as follows: $i)$ Prepare a three-qubit state in $\ket{\psi}\ket{\Phi^+}$; $ii)$ apply $\M{H}_1\M{{CX}}_{1,2}$; $iii)$ measure the first and the second qubit in the $\{ \ket{0}, \ket{1} \}$-basis; $iv)$ based on the results of the measurements $(m_1, m_2) \in \{0, 1\}^2$, apply $\M{X}_3^{m_2}\M{Z}_3^{m_1}$ to obtain $\ket{\psi}$ on the third qubit. 
Since the first operation acts only on the first two qubits, it is possible to transfer an unknown quantum state at distance having a pre-shared \ac{EPR} pair and transmitting two bits of information over a classical channel.

\subsection{Quantum Distillation Protocols}
\label{subsec:QPurification}
In order to reliably teleport a quantum state it is assumed to have a perfect pre-shared \ac{EPR} pair $\ket{\Phi^+}$. However, imperfect local operations and entanglement generation at distance induce errors on the pre-shared pair, which lead to mixed states. 
We describe the mixed state with the density matrix
\begin{align}
\label{eq:GenMixedState}
\M{\rho} &= A \ket{\Phi^+}\bra{\Phi^+} + B \ket{\Psi^-}\bra{\Psi^-} \notag \\
&+ C \ket{\Psi^+}\bra{\Psi^+} + D \ket{\Phi^-}\bra{\Phi^-}
\end{align}
where the coefficients are real, normalized $A + B + C + D = 1$, and defined on the interval $[0, 1]$. The coefficient $A$ is the entangled pair fidelity.
An important mixed state is the maximally mixed one, referred to in the following as a Werner state \cite{Wer89:WernerState, Ben96:purification, Zha02:WernerPrep}, having density matrix
\begin{align}
\label{eq:Werner}
\M{\rho} &= \mathcal{F} \ket{\Phi^+}\bra{\Phi^+} \notag \\
&+ \frac{1-\mathcal{F}}{3} \Big[ \ket{\Psi^-}\bra{\Psi^-} + \ket{\Psi^+}\bra{\Psi^+} + \ket{\Phi^-}\bra{\Phi^-}\Big]\,.
\end{align}

\begin{figure}[t]
    \centering
    \includegraphics[width = 0.6\columnwidth]{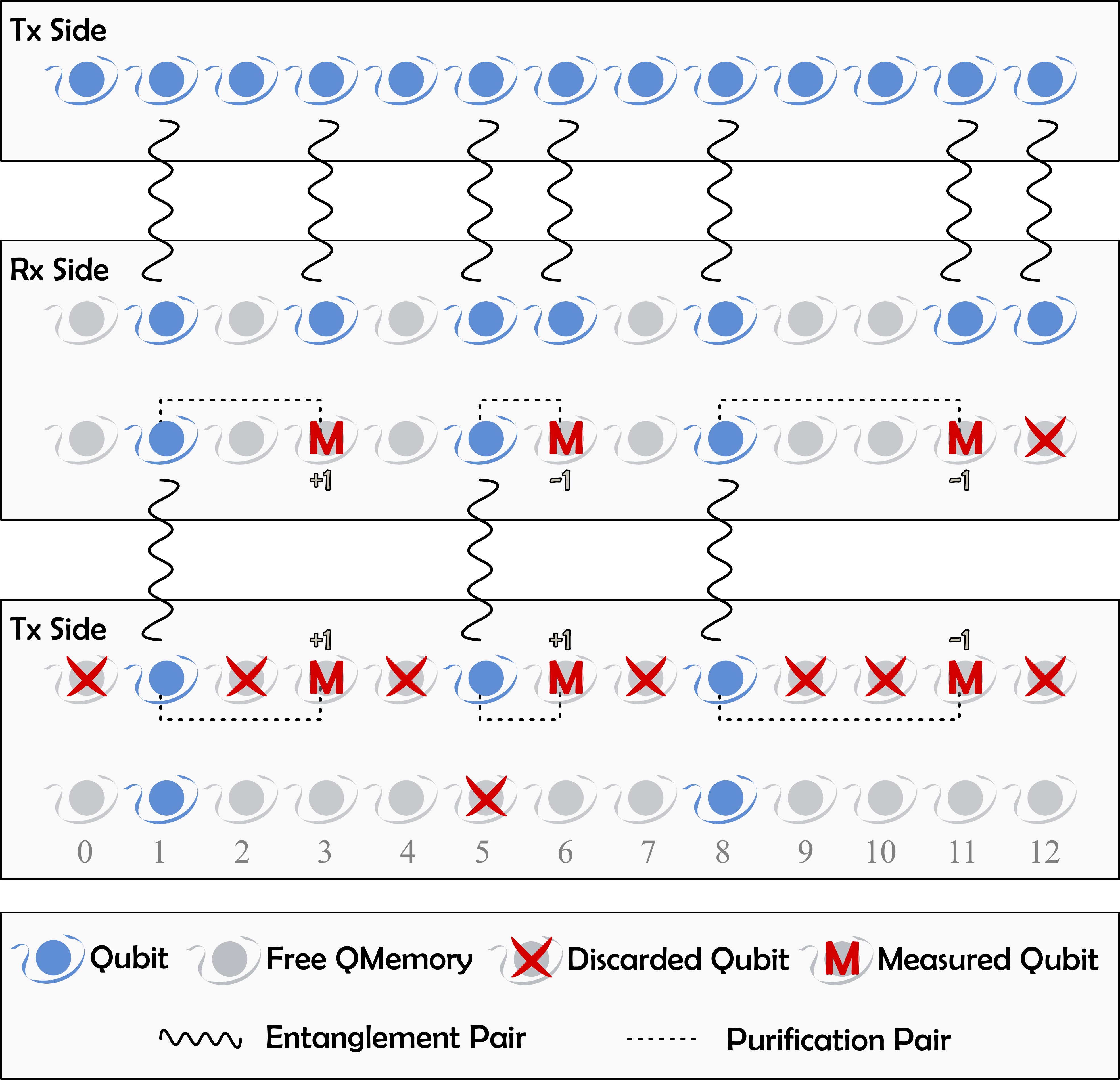}
    \caption{Pictorial representation of raw entanglement distribution and one step of entanglement purification. In this example, the transmitter attempts to share $13$ qubits, where each of them is part of a different \ac{EPR} pair. The receiver successfully detects $7$ qubits among the $13$ transmitted ones. It groups $3$ pairs for purification, performs it, and discards the unpaired one. Finally, it sends the information about which qubit has to be kept and measurements to the original transmitter. Concluding the purification procedure in this example, we have $2$ \ac{EPR} pairs in position $1$ and $8$ of the user's respective quantum memories.}
    \label{fig:QuantumPurificationPictorial}
\end{figure}

Several techniques have been developed to increase the fidelity of entangled states for teleportation. Here we consider the distillation algorithm presented in \cite{Deu96:purification} which improves \cite{Ben96:purification}.
The distillation protocol can be summarized as follows \cite{Das21:IBMQPuri}: $i)$ share two \ac{EPR} pairs (i.e., first pair: qubits $1$ and $2$, second pair: qubits $3$ and $4$), both in state \eqref{eq:GenMixedState}; $ii)$ apply the local rotations $\M{R}_{x_1}(\pi/2) \M{R}_{x_2}(-\pi/2) \M{R}_{x_3}(\pi/2) \M{R}_{x_4}(-\pi/2)$; $iii)$ apply the local CNOTs $\M{{CX}}_{1,3}\M{{CX}}_{2,4}$; $iv)$ measure qubits $3$ and $4$ and share via classical messages the measurement information; $v)$ if the measurements agree (i.e., both $0$s or $1$s) keep the final pair, otherwise discard it.
Through the paper, we consider that errors in quantum measurements, quantum local operations, and classical communications can be neglected. In \cite{Dur99:NoisyPurif,Dur07:PuriAndApplications,ChiSim22:LearningPurifNoisyClasMsg} some of these aspects have been accounted for, although not adopting \cite{Deu96:purification} which is our focus in this work. 
Since the procedure from $ii)$ to $iv)$ uses only local operations and measurements, these steps can be efficiently pipelined to minimize the latency of the communication protocol \cite{Van08:Qrepeater}. A pictorial representation of the distillation procedure is provided in Fig.~\ref{fig:QuantumPurificationPictorial}.
The resulting mixed state, considering equally distributed \ac{EPR} pairs (also referred as symmetric distillation), is described by
\begin{equation}
\label{eq:DeautschProbEvol}
\begin{aligned}
A_{i+1} &= (A_{i}^2 + B_i^2) \, N_i^{-1}\\
B_{i+1} &= 2 C_{i} D_i N_i^{-1}\\
C_{i+1} &= (C_{i}^2 + D_i^2) \, N_i^{-1}\\
D_{i+1} &= 2 A_{i} B_i N_i^{-1}\\
\end{aligned}
\end{equation}
where the subscripts indicate the number of distillation steps and $N_i = (A_{i} + B_i)^2 + (C_{i} + D_i)^2$ is a normalization factor. 
The convergence of \eqref{eq:DeautschProbEvol} to $A_i \to 1$ when $i\to \infty$ was proven in \cite{Mac98:ProofPurification}. 
The parameter $N_i$ also represents the probability that the $i$-th distillation step succeeds. 
Due to this random behavior in the generation, we have that initially the transmitter does not know which qubit is effectively useful. For this reason, it has to store them until the receiver confirms the entanglement generation (see the pictorial example in Fig.~\ref{fig:QuantumPurificationPictorial}). 
As remarked in \cite{Deu96:purification}, the twirling performed in \cite{Ben96:purification} (i.e., an operation which uniformly distributes the error states resulting in $B = C = D$), has the effect to slow the convergence of $A_i \to 1$.
For this reason, in the following we consider the initial state to be a Werner one to show how ad-hoc coding can improve this critical scenario. 

In recent years, several distillation protocols have been proposed \cite{KraAlbJia2019:OptPurification, Goo2023:nkPurification}. However, we stick with \cite{Deu96:purification} for simplicity in the analytical treatment.
On the other hand, it is possible to use the proposed techniques to search for asymmetry in other distillation protocols, and then use ad-hoc error correction as we will show in the following.
We will touch upon the analysis of other protocols in Section~\ref{sec:NumRes}.

\subsection{Quantum Error Correction}
\label{subsec:QEC}
We indicate with $[[n,k,d]]$ a \ac{QECC} that encodes $k$ information qubits into a codeword of $n$ qubits, having distance $d$. This code is able to correct all patterns of up to $t = \lfloor(d-1)/2 \rfloor$ errors. 
We consider stabilizer codes $\mathcal{C}$ generated by $n-k$ independent and commuting operators $\M{G}_i \in \mathcal{G}_n$, called generators, where $\mathcal{G}_n$ is the Pauli group on $n$ qubits \cite{Got:09,NieChu:10}.
The code $\mathcal{C}$ is the set of quantum states $\ket{\psi}$ satisfying $\M{G}_i \ket{\psi}=\ket{\psi},\, i=1, 2, \ldots, n-k$. 
These error-correcting codes preserve their information by means of measurements of extra qubits (usually referred to as ancillas) which have been properly entangled with the codeword.
For the sake of clarity, assume a codeword $\ket{\psi} \in \mathcal{C}$ is affected by a channel error. 
Measuring the codeword according to the stabilizers $\M{G}_i$ with the aid of ancilla qubits, the error collapses on a discrete set of possibilities represented by combinations of Pauli operators $\M{E}\in \mathcal{G}_n$ \cite{Got:09}. 
For example, an error represented by a rotation $\M{R}_x(\theta)$ collapses into $\M{I}$ or $\M{X}$ with probability $\cos^2(\theta/2)$ and $\sin^2(\theta/2)$, respectively when the ancillae are measured according to the stabilizer. For this reason, quantum channels for stabilizer codes are described by specifying the single Pauli error probabilities. 
The measurement procedure over the ancilla qubits results in a quantum error syndrome $\Syndrome(\M{E})=(s_1, s_2, \ldots,s_{n-k})$, with each $s_i =0$ or $1$ depending on $\M{E}$ commuting or anticommuting with $\M{G}_i$ \cite{Got:09}. 
A maximum likelihood decoder will then infer the most probable error $\M{\hat{E}}\in \mathcal{G}_n$ compatible with the measured syndrome. On the other hand, considering decoders correcting codewords with up to $t = \lfloor (d-1)/2 \rfloor$ errors (i.e., the guaranteed error correction capability of a bounded distance decoder) we have that the logical qubit error probability is
\begin{equation}
\label{eq:PeGen}
\rho_\mathrm{L} = 1 - \sum_{j = 0}^{t} \binom{n}{j} \rho^j \, (1-\rho)^{n-j}
\end{equation}
where $\rho$ is the physical qubit error probability. This can be used to upper bound the error probability of the maximum likelihood decoder, while, for codes with negligible degeneracy effects, is also a tight approximation \cite{ValFor23:Surface}.
%
In the following we also consider $[[n, k]]$ asymmetric codes able to correct up to $e_\mathrm{g}$ generic errors (i.e., $\M{X}$, $\M{Y}$, $\M{Z}$, or none) plus up to $e_\mathrm{Z}$ Pauli $\M{Z}$ errors, and no others \cite{ChiVal:20a}. In the presence of asymmetry in the error probabilities of a quantum channel, these codes can obtain improvement in performance and code length efficiency.
In this setting, we define $p_\mathrm{X}$, $p_\mathrm{Y}$ and $p_\mathrm{Z}$ as the probability to have an error of type $\M{X}$, $\M{Y}$, and $\M{Z}$, respectively. %
Then, by weighting each pattern with the corresponding probability of occurrence, the expression \eqref{eq:PeGen} is generalized by \cite{ChiVal:20a}
\begin{align}
\label{eq:PeGen2}
\rho_\mathrm{L} 
&=1 - \sum_{j = 0}^{e_\mathrm{g}+e_\mathrm{Z}} \binom{n}{j}(1-\rho)^{n-j}\sum_{i = (j-e_\mathrm{g})^+}^{j}\binom{j}{i} \, p_\mathrm{Z}^i \, \left(\rho-p_\mathrm{Z}\right)^{j-i}
\end{align}
where $(x)^+=\max\{x,0\}$ and $\rho = p_\mathrm{X} + p_\mathrm{Y} + p_\mathrm{Z}$.

\section{Ad-hoc Coding over Teleportation Channels}
\label{sec:mainContribution}
\subsection{Quantum Teleportation Channel: Definition}
\label{subsec:QTPChannel}

Among quantum channels, a subclass of them is described by Pauli errors, for example, due to the error collapse briefly described in Section~\ref{subsec:QEC}. 
For such channels, the qubits passing through them can experience a Pauli error $\M{X}$, $\M{Y}$, and $\M{Z}$ or none.
Respectively, they occur with probabilities $p_\mathrm{X}$, $p_\mathrm{Y}$, and $p_\mathrm{Z}$, while no errors occur with probability $1-\rho$, with $\rho = p_\mathrm{X} + p_\mathrm{Y} + p_\mathrm{Z}$. In particular, some well-known cases are represented by: $i)$ the bit-flip (phase-flip) channel where a qubit can experience only a $\M{X}$ ($\M{Z}$) error with probability $p_\mathrm{X}$ ($p_\mathrm{Z}$); $ii)$ the depolarizing channel where $p_\mathrm{X} = p_\mathrm{Y} = p_\mathrm{Z}$ \cite{NieChu:10}; $iii)$ the asymmetric channel characterized by asymmetry parameter $\mathcal{A} = p_\mathrm{Z}/p_\mathrm{X}$ where $p_\mathrm{X} = p_\mathrm{Y}$ and $p_\mathrm{Z} = \mathcal{A}\,\rho/(\mathcal{A}+2)$ \cite{Sar:2009, ChiVal:20a}; $iv)$ the independent $\M{X}\M{Z}$ channel where each qubit pass through a concatenation of a bit-flip and a phase-flip channel \cite{Mac04:QLDPC}. For asymmetric channels, \eqref{eq:PeGen2} can be expressed as a function of $\rho$ and $\mathcal{A}$ instead of $\rho$ and $p_\mathrm{Z}$. Generally, we define the equivalent asymmetric parameter $\mathcal{A}_\mathrm{eq} = 2\,p_\mathrm{Z} / (p_\mathrm{X} + p_\mathrm{Y})$, resulting in a logical qubit error probability
\begin{align}
\label{eq:PeGen3}
\rho_\mathrm{L} =1 - \sum_{j = 0}^{e_\mathrm{g}+e_\mathrm{Z}} \binom{n}{j} \, &(1-\rho)^{n-j} \rho^j 
\left(\frac{2}{\mathcal{A}_\mathrm{eq} + 2}\right)^j \notag \\
&\times \sum_{i = (j-e_\mathrm{g})^+}^{j}\binom{j}{i} \, \left(\frac{\mathcal{A}_\mathrm{eq}}{2}\right)^{i}
\end{align}
when asymmetric codes are adopted. On the other hand, symmetric codes whose performance is described by \eqref{eq:PeGen} are not affected by any error imbalance since they act on generic errors.
Since the asymmetric channel is a particular case of the defined equivalent asymmetric channel, in the following we use the $\mathcal{A}_\mathrm{eq}$ parameter dropping the notation equivalent for the sake of simplicity.

\begin{figure}[t]
    \centering
    \includegraphics[width = 0.7\columnwidth]{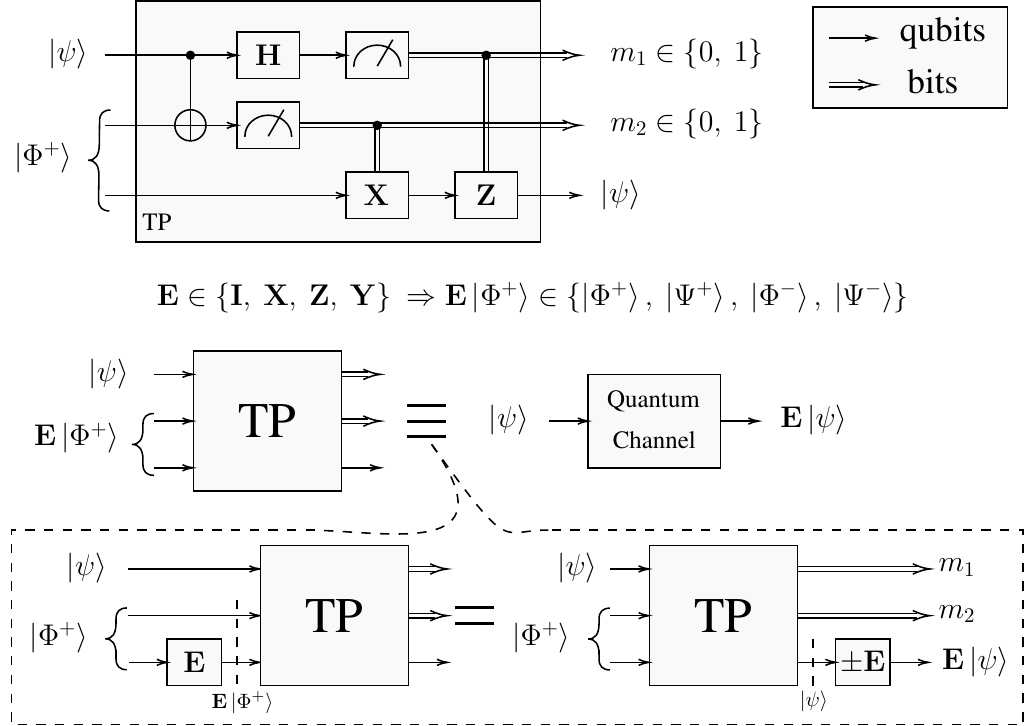}
    \caption{Equivalence between teleportation protocol using noisy \ac{EPR} pair and a quantum communication channel based on Pauli errors.}
    \label{fig:QTPCh}
\end{figure}

Due to imperfections in \ac{EPR} pairs, the quantum teleportation protocol can be interpreted as a quantum communication channel \cite{Ben96:purification}. 
In fact, considering an information state $\ket{\psi}$ to be teleported and the \ac{EPR} pair $\ket{\Psi^+}$ instead of $\ket{\Phi^+}$ as described in Section~\ref{sec:preliminary}, at the end of the teleportation algorithm we obtain $\M{X} \ket{\psi}$.
In general, considering a mixture of \ac{EPR} pairs as in \eqref{eq:GenMixedState}, we can describe the quantum teleportation with a surrogate channel having $p_\mathrm{X} = C$, $p_\mathrm{Y} = B$, and $p_\mathrm{Z} = D$ (see Fig.~\ref{fig:QTPCh}).  
We point out that in particular, assuming a Werner state as the shared pair, teleportation is equivalent to transmission over a depolarizing channel.

This observation, despite its simplicity, is the bridge to connect quantum distillation and \ac{QEC} in the direction of reliable low-latency communications, adaptive \ac{QEC} for quantum networks, and mid-generation error management schemes.

\subsection{Quantum Teleportation Channels: Analysis}
\label{subsec:AdHocCodingTP}
In this subsection, we report an analysis of the distillation protocol proposed in \cite{Deu96:purification}, described in Section~\ref{subsec:QPurification}.
Our aim is to show that distillation protocols tends to generate asymmetries in the output entangled pair, which can be exploited by asymmetric error correction schemes.
In particular, we consider shared pairs in the Werner state \eqref{eq:Werner} with an initial fidelity $\mathcal{F}_0$.
The following analysis can be conducted to search for asymmetries in other distillation protocols or under different initial conditions. Yet, the analytical treatment could not be always feasible.

In our setting, we have that after one step of distillation the state evolves into
\begin{equation}
\label{eq:OneStepPurWerner}
\begin{aligned}
A_{1} &\propto \mathcal{F}_0^2 + (1-\mathcal{F}_0)^2 / 9\\
B_{1} &\propto 2 (1-\mathcal{F}_0)^2 / 9\\
C_{1} &\propto 2 (1-\mathcal{F}_0)^2 / 9\\
D_{1} &\propto 2 \mathcal{F}_0 \, (1-\mathcal{F}_0) / 3\\
\end{aligned}
\end{equation}
where we have omitted the common normalization factor. We observe that the equivalent quantum teleportation channel evolved from the depolarizing channel to the asymmetric channel described in Section~\ref{subsec:QTPChannel}. To describe such a channel it is sufficient to give the error probability $\rho$ and the equivalent asymmetry parameter $\mathcal{A}_\mathrm{eq}$. In our case we have 
\begin{align}
\label{eq:AsymPara}
\mathcal{A}_{\mathrm{eq},1} &\triangleq \frac{2\,D_1}{B_1+C_1} = 3 \left(\frac{1}{\rho_0}-1\right) \\
\label{eq:ErrProbOneStepPur}
\rho_1 &\triangleq B_1+C_1+D_1 = 2 \rho_0 \frac{3-\rho_0}{9 - 12\,\rho_0 + 8\,\rho_0^2}
\end{align}
where $\rho_0 = 1 - \mathcal{F}_0$ is the initial error probability. From \eqref{eq:ErrProbOneStepPur} we note that, even considering $\rho_0 \ll 1$, the error probability tends to $2\,\rho_0/3$. 
This limited improvement is the cause of the algorithm in \cite{Ben96:purification} requiring more distillation steps than \cite{Deu96:purification} in order to acquire the same target fidelity. 
Considering symmetric \ac{QEC}, the improvement given by one step of distillation is likely not worth it compared to the cost in raw \ac{EPR} pairs. However, in asymmetric \ac{QEC} the performance is affected both by the error and asymmetry parameters, making them appealing for this scenario. 
In fact, in \eqref{eq:AsymPara} we can observe the effectiveness of the first step of distillation in increasing the asymmetry of the channel from $\mathcal{A}_\mathrm{eq} = 1$ to $\mathcal{A}_\mathrm{eq} \approx 3/\rho_0$.

\begin{figure}[t]
  \centering
  \includegraphics[width = 0.7\columnwidth]{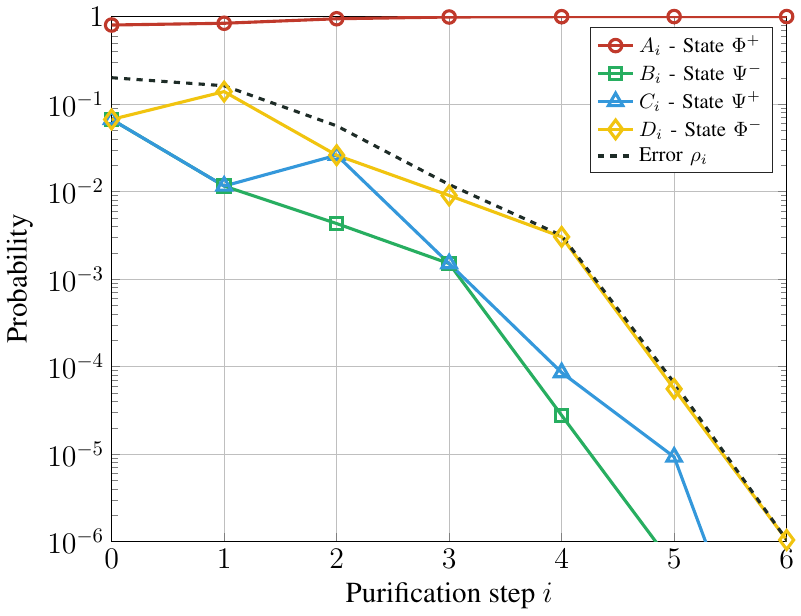}
  \caption{\justifying Evolution of the probability distribution of a \ac{EPR} pairs mixture described by \eqref{eq:GenMixedState} due to distillation algorithm \eqref{eq:DeautschProbEvol}. The initial state is a Werner state with fidelity $\mathcal{F}_0 = 0.8$.}
  \label{fig:WernerStateEvolution}
\end{figure}

\begin{figure}[t]
  \centering
  \includegraphics[width = 0.7\columnwidth]{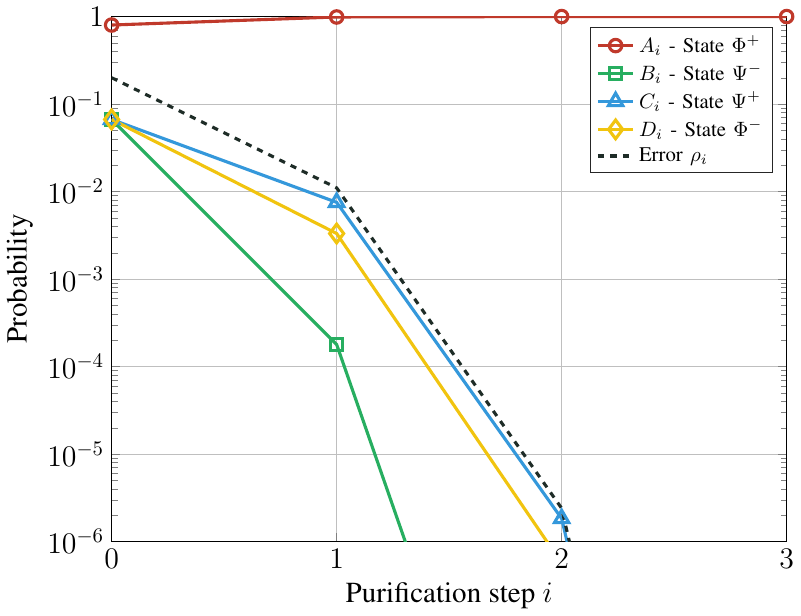}
  \caption{\justifying Evolution of the probability distribution of a \ac{EPR} pairs mixture described by \eqref{eq:GenMixedState} due to the STRINGENT distillation algorithm~\cite{Nickerson2013}. The initial state is a Werner state with fidelity $\mathcal{F}_0 = 0.8$.}
  \label{fig:WernerEvolutionStringent}
\end{figure}

In Fig.~\ref{fig:WernerStateEvolution} we report the state evolution of the \ac{EPR} pairs distilled by \eqref{eq:DeautschProbEvol} and assuming to start from the Werner state with $\mathcal{F}_0 = 0.8$. 
As shown in \eqref{eq:OneStepPurWerner}, and as depicted in Fig.~\ref{fig:WernerStateEvolution}, the  quantum teleportation channel after one step of distillation becomes asymmetric in the conventional sense (i.e., $p_\mathrm{X} = p_\mathrm{Y}$). 
More in general, this happens whenever the previous state has $C_{i-1} = D_{i-1}$. In fact, having a state with parameter $\{A, B, C, D\}$ equal to $\{A, B, k, k\}$, by applying \eqref{eq:DeautschProbEvol} we end up to a state with $\{A^2+B^2, 2\,k^2, 2\,k^2, 2\,AB\}$ where the common normalization factor has been omitted. 
Observing the distillation procedure only between $i=0$ and $i=4$, it seems that having an asymmetric channel at step $i$ provides a state with $C_{i+1} = D_{i+1}$ at step $i+1$. However, this is not true in general as has been reported in the plot evolving the state from step $i=3$ to $i=4$ (or by checking \eqref{eq:DeautschProbEvol}). The fact that we have $C_2 = D_2$ is a consequence of the Werner state assumption $B_0 = C_0 = D_0$ and not due to $B_1 = C_1$. Noted that this initial behavior is not recursive we observe that the teleportation channel tends to be asymmetric in the sense that $D_i \gg B_i+C_i$ for $i > i_\mathrm{min}$ where $i_\mathrm{min}$ depends on the initial fidelity $1 - \rho_0$. 
We conclude that asymmetries arise in the distilled pairs.
Hence, exploiting this feature of distillation protocols could play a key role to improve the performance of quantum communication.

As a comparison to newer distillation protocols, we show in Figure~\ref{fig:WernerEvolutionStringent} the same setup as in Figure~\ref{fig:WernerStateEvolution}, but using the STRINGENT protocol~\cite{Nickerson2013} rather than the Deutsch et al. distillation in~\eqref{eq:DeautschProbEvol}.
While we do see a larger improvement in fidelity in each step, one should note that the STRINGENT circuit is more complicated and requires a large amount of gates and measurements compared to the distillation in~\eqref{eq:DeautschProbEvol}.
This also shows in the success probabilities. For the initial fidelity $\mathcal{F}_0=0.8$ considered here, the first step of distillation using STRINGENT has 
${\sim}5.8\%$ success probability, while increasing the fidelity to ${\sim}0.99$.
On the other hand, the three first steps of the older distillation protocol by Deutsch et al. have a combined success probability of ${\sim}51.6\%$ while still delivering a fidelity of ${\sim}0.99$.
Clearly, a lower success probability implies that more distillation attempts are necessary to obtain the same number of distilled pairs. Thus, one needs to either perform more distillations in parallel (increasing the requirements on quantum memory and processing power), or more distillations sequentially (increasing the latency). 
Neither option is desirable in our setting, for which reason we will restrict ourselves to the distillation procedure summarized in~\eqref{eq:DeautschProbEvol}.

\begin{figure}[t]
    \centering
    \includegraphics[width = 0.7\columnwidth]{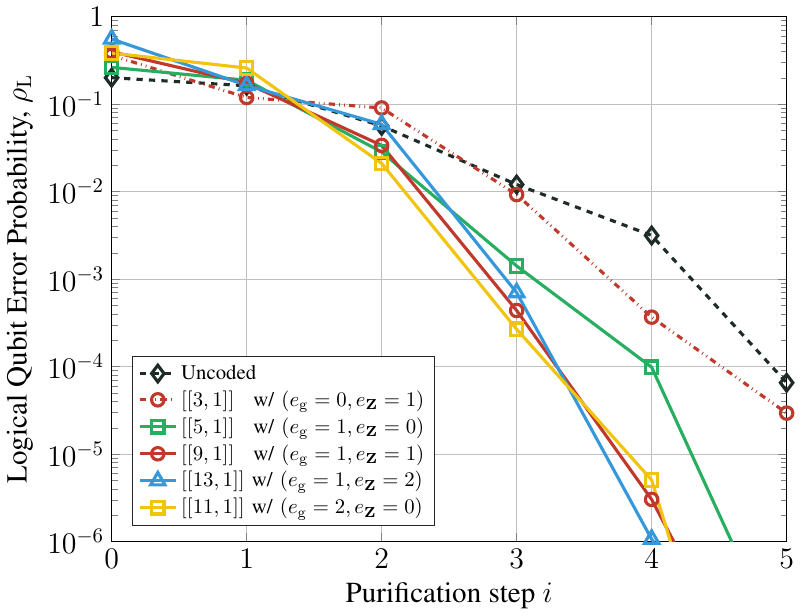}
    \caption{\justifying Logical qubit error probability as a function of distillation steps for different coding schemes. We consider bounded distance decoding for $[[n,k]]$ codes with $(e_\mathrm{g}, e_\mathrm{Z})$ error correction capability as reported in \eqref{eq:PeGen3}. The initial state is a Werner one with fidelity $\mathcal{F}_0 = 0.8$.}
    \label{fig:LatencyGain}
\end{figure}

Ad-hoc \ac{QEC} represents one possible way to exploit the particular channel configurations which are evolving due to distillation.
To this aim, we report in Fig.~\ref{fig:LatencyGain} the performance, in terms of logical error probability (i.e., one minus the reliability), of some $[[n,k]]$ codes with $(e_\mathrm{g}, e_\mathrm{Z})$ error correction capability, varying the number of distillation steps $i$. 
In general, for codes with $e_\mathrm{g} >0$ we have that after a distillation step $\ell$ the coded scheme starts to outperform the uncoded one and it continues to outperform the uncoded for each $i > \ell$. 
In particular, $\ell$ depends on the code and the initial error probability $\rho_0$, which is fixed to $\rho_0 = 0.2$ in the plot.
Reading the plot horizontally, it is also possible to obtain some insights in terms of latency. In fact, when targeting a reliability $1-\rho_\mathrm{L}^*$, we have that the uncoded scheme requires more distillation steps, which is in some situations equivalent to saying that it requires more classical protocol messages. 
Finally, we want to emphasize that, observing only the symmetric codes, we have that larger codes are always better after a certain threshold $\ell$.
In other words, after a certain threshold, there is an order on the $\rho_\mathrm{L}$ of the codes which is preserved, due to the fact that the distillation protocol converges to fidelity one \cite{Mac98:ProofPurification}, and larger codes have larger $e_\mathrm{g}$.
On the other hand, when asymmetric codes are considered, the channel asymmetry $\mathcal{A}_{\mathrm{eq},i}$ becomes an important design parameter alongside $\rho_i$. 
Hence, the code must be chosen carefully depending on the particular quantum teleportation channel.

\subsection{Low Latency Protocols: Single Quantum Link}
\label{subsec:SingleLink}

In this section, we address on-demand single quantum link teleportation protocols having low latency constrains. 
In particular, we aim at minimizing the number of message exchanges necessary to establish reliable communication.
Let us begin with a forward communication protocol between two nodes.

\begin{figure}[t]
    \centering
    \includegraphics[width = 0.6\columnwidth]{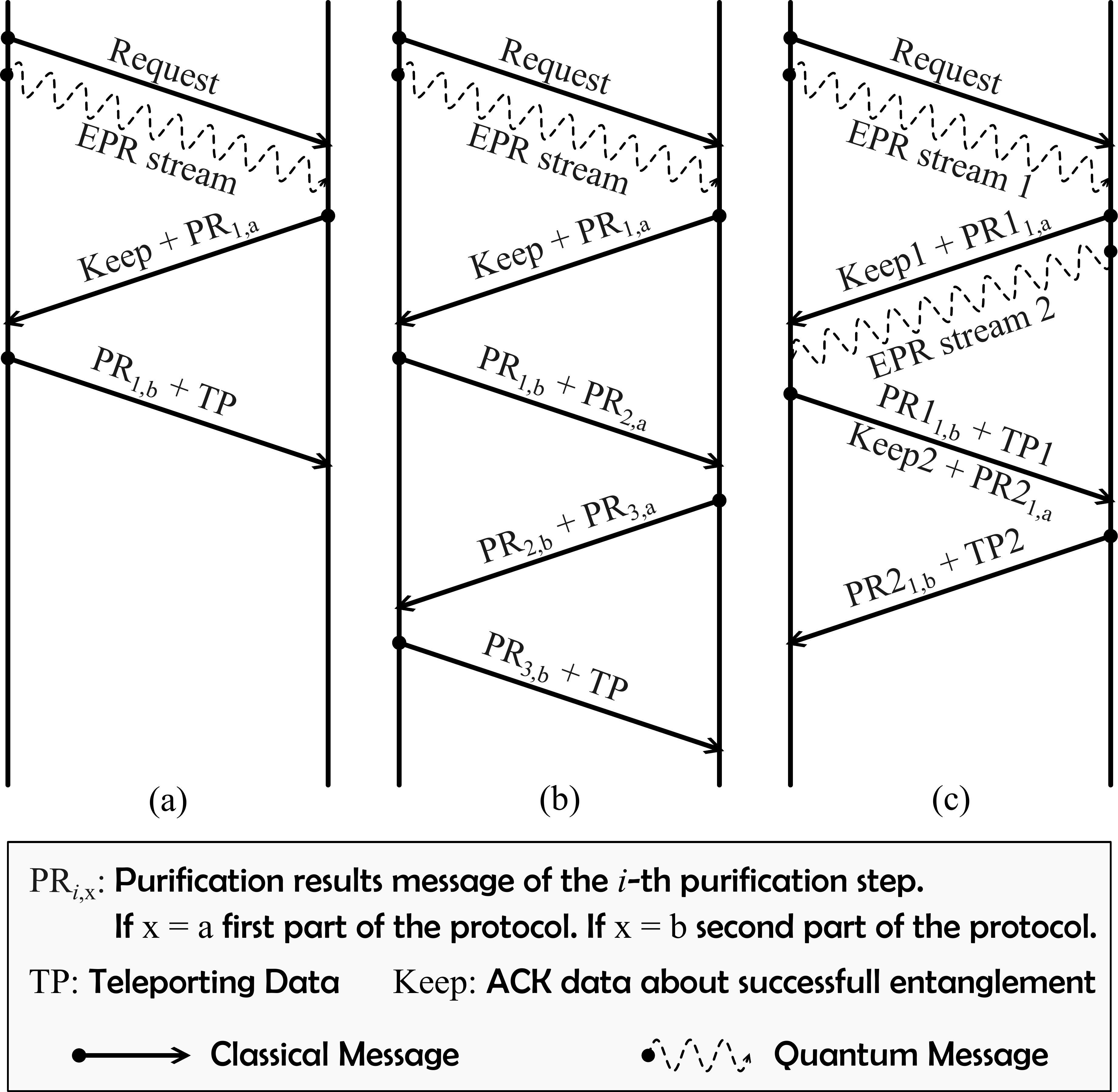}
    \caption{\justifying Protocol messaging for \ac{EPR} pair creation, distillation and teleportation. Message exchange for: (a) single distillation; (b) three distillation; (c) back-and-forward protocol with single distillation. }
    \label{fig:ProtocolMessages}
\end{figure}

As discussed in \cite{Van08:Qrepeater} and illustrated in Fig.~\ref{fig:ProtocolMessages}a, it is possible to create a quantum channel by the mean of distilled \acp{EPR} (one distillation step) with a three steps procedure. 
Firstly, the node which desires to communicate something generates $M$ \ac{EPR} pairs and sends a request message followed by the qubits for entanglement distribution among the two nodes \cite{Cir97:EntanGen, Chi05:EntanGen, Loo06:EntaGen, Uph16:EntaGen, Hu21:EntaGen}.
This heralded entanglement generation is classified as ``at source'' generation \cite{Cac20:QTPforQI,irtf-qirg-principles-11}.
The receiver, if not busy, stores into its quantum memory the qubits whose entanglement is guaranteed by a detector \cite{Cir97:EntanGen,Bar05:DetectorEntanGen}. Then, it performs the local operations required for the first step of distillation as reported in Section~\ref{subsec:QPurification}. 
Defining $p$ as the probability to receive a qubit and that it is correctly entangled, the number of received qubits $n_0$ is distributed according to a binomial distribution with $M$ trials and success probability $p$ (sometimes also referred to as the emission probability).

Secondly, it transmits the information regarding which initial \ac{EPR} pair has been successfully received (usually referred to as a \emph{keep} message), which pairs have been selected for distillation, and its partial distillation results obtained by the measurement step. 
Upon receiving this information, the user who initially asks for the channel can apply the distillation procedure on the appropriate pairs and based on the results understand which pair have been successfully distilled. 
At this point, the number of \ac{EPR} pairs $n_1$ after one step of distillation is given again by a binomial distribution with $\lfloor n_0 / 2 \rfloor$ trials and success probability $N_1 = (A_1 + B_1)^2 + (C_1 + D_1)^2$ due to the distillation described in \eqref{eq:DeautschProbEvol}.
Referring to the example shown in Fig.~\ref{fig:QuantumPurificationPictorial}, we have that starting with $M=13$ pairs, the receiver detects $n_0 = 7$ qubits, and at the end of one distillation step we end up with $n_1 = 2$ distilled \ac{EPR} pairs.

Lastly, the user can either concatenate another distillation step, concatenate the quantum information teleportation, or simply end the distillation, sending the measurement results. 

To nest a second distillation step the transmitter has to repeat the same operations performed by the receiver at step two. 
Consequently, it appends to the first distillation measurement data (which acts as a keep message at this point), the new measurements, and corollary information.
As an example, we report in Fig.~\ref{fig:ProtocolMessages}b a protocol with three nested distillations.
Due to the fact that we are targeting low latency protocols, we consider directly teleporting quantum data. 
This also represents the protocol with the lowest latency since entanglement generation is a sporadic event, making the \emph{keep} message mandatory. 
To be precise, without \ac{EPR} pair confirmation, the link fidelity would be lower bounded by entanglement generation probability (usually very low for long-distance generation), which is reflected in an unusable link. 
Hence, transmitting this \emph{keep} message, one distillation step can always be performed without any significant cost in latency.

To increase the fidelity we propose a hybrid scheme using both distillation and \acf{QEC}. 
In fact, instead of using the teleportation protocol on the quantum data, it is possible to teleport a quantum codeword that represents the encoded data. 
As usual, applying \ac{QEC} we aim to improve the reliability of the system at the cost of transmission rate. 
In particular, adopting an $[[n, k]]$ code we need $n$ \ac{EPR} pairs to teleport $k$ information qubit. 
Since in mid-generation it could be difficult to construct long and complex \acp{QECC}, which are the target for third-generation error management schemes \cite{irtf-qirg-principles-11}, in the following we consider short codes \cite{Sho:95, Laf:96, SteAnd:96, ChiVal:20a}. 

In \eqref{eq:AsymPara} we have observed that starting from a Werner state, we gain asymmetry in the equivalent quantum channel after one distillation step.
Taking this into account, as shown by the generalized quantum Hamming bound \cite{ChiVal:20a}, we can design ad-hoc codes which either use fewer qubits to achieve the same performance or vice versa.
In this asymmetric scenario, also very simple codes such as the $[[n, 1]]$ with $e_\mathrm{g} = 0$ and $e_\mathrm{Z} = \lfloor n/2 \rfloor$ (i.e., repetition codes), can be of practical interest. 
This is motivated by the encoder and decoder simplicity and the performance boost they can provide in the presence of strong asymmetry.
More generally, adopting $[[n, k]]$ with error correction capability $(e_\mathrm{g}, e_\mathrm{Z})$ in this hybrid communication protocol we will show in Section~\ref{sec:NumRes} the overall achievable fidelity improvements.

Finally, we discuss the possibility to have a reliable back-and-forward communication protocol. 
This protocol, depicted in Fig.~\ref{fig:ProtocolMessages}c, is intended for applications in which a node has to send a qubit to a neighbor for multi-qubit processing. 
This operation (e.g., a CNOT) is performed between the neighbor qubits and the sent qubit, consequently, it is sent back to its owner.
Some application examples are represented by multi-core quantum computing \cite{Bal22:MultiCoreQComp}, construction of quantum graph state \cite{Hei04:MultipartyGraphStates,Hei06:GraphStates, Mei19:DistribGraphStates} over networks, and in general, all applications where it is useful to extend the concept of local operations to neighbors.
In comparison with the conventional protocol in Fig.~\ref{fig:ProtocolMessages}a having only the \emph{forward-link} created using the first \ac{EPR} stream, during the second transmission a second \ac{EPR} stream is concatenated to create the \emph{back-link}.
As usual, after three transmissions the receiver holds the transmitter qubit and can perform its processing. When the processing is done, using the \emph{back-link} it is possible to send back the qubit with just one classical transmission. 
Increasing the stream size and the initial quantum memory $M$ accordingly, it is possible to parallelize this procedure to transmit a packet of qubits. 
This can be done using $[[n, k]]$ codewords with $k > 1$ or concatenation of $[[n, 1]]$ codewords.

\emph{Note:} a double-size quantum memory is not required for this link communication. This is due to the fact that the receiver can allocate its new \ac{EPR} pair qubits (\ac{EPR} stream $2$) in memories that have non-generated entangled pairs, measured qubits, and discarded qubits. For example, in Fig.~\ref{fig:QuantumPurificationPictorial}, \ac{EPR} stream $1$ is composed of $13$ qubits, while \ac{EPR} stream $2$ might be composed by up to $10$ qubits. For this reason, the initial memory $M$ may increase, but it is not necessary to double it.

\subsection{Low Latency Protocols: Quantum Networks}
\label{subsec:MultiLinks}

\begin{figure}[t]
    \centering
    \includegraphics[width = 0.99\columnwidth]{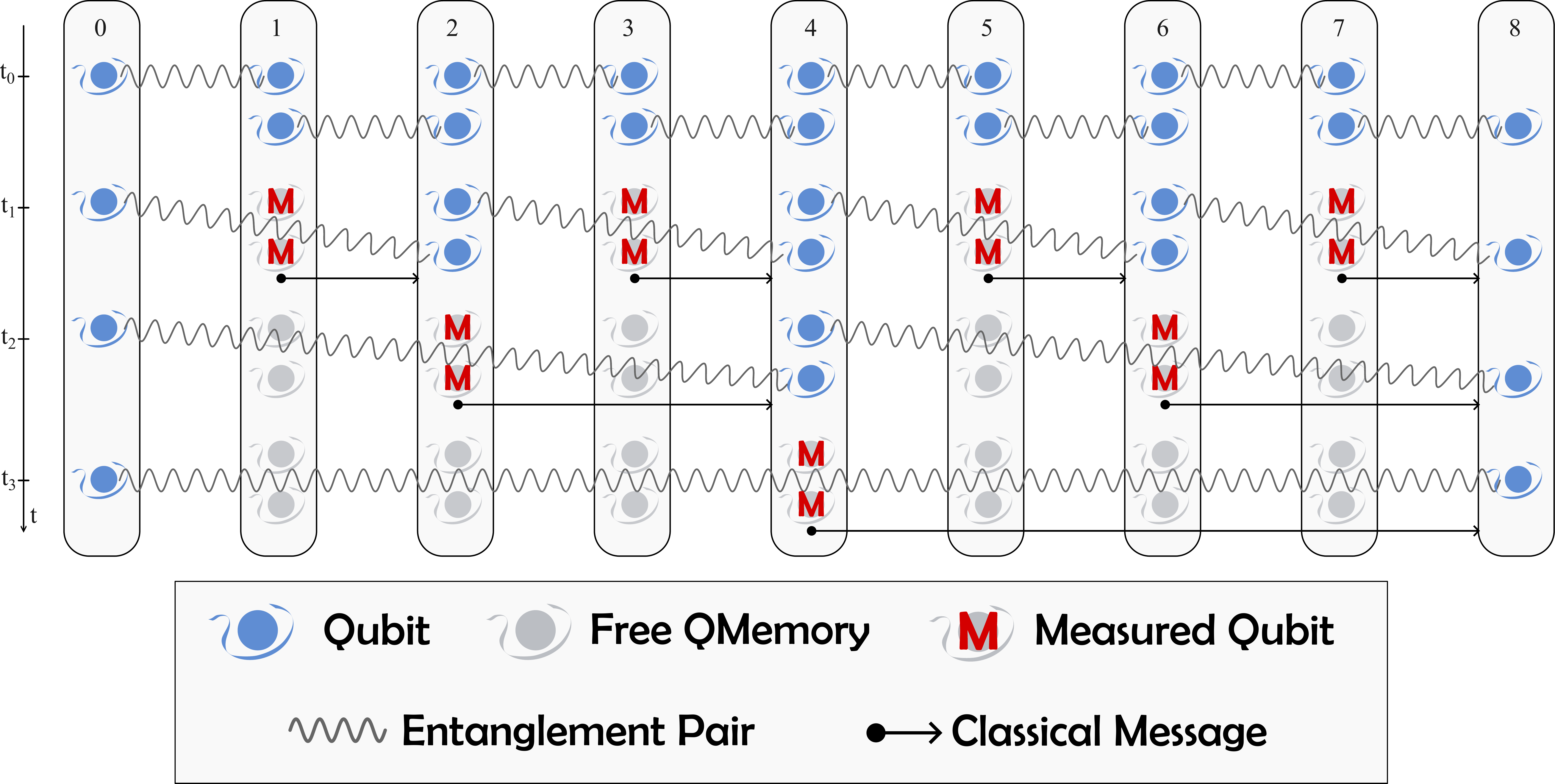}
    \caption{\justifying Nested entanglement swapping procedure to share \ac{EPR} pairs between far away nodes. In the example, we have a network with $8$ hops which requires $3$ swapping steps.}
    \label{fig:ExampleEntanglementSwapping}
\end{figure}

The entanglement distribution over quantum networks is achieved using the entanglement swapping protocol \cite{irtf-qirg-principles-11}. This protocol is the conventional quantum teleportation which, instead of transmitting one qubit of information, transmits one qubit of an entangled pair \cite{BriDur98:QSwapAndPur}. 
For the sake of clarity, let's consider this example. Three users $u_1$, $u_2$, and $u_3$ share two \ac{EPR} pairs $\ket{\Phi^+}_{12}$ and $\ket{\Phi^+}_{23}$, where the subscripts indicate which users have the qubits. User $2$ can teleport its qubit of the pair $\ket{\Phi^+}_{12}$ to $u_3$ using $\ket{\Phi^+}_{23}$ as described in Section~\ref{sec:preliminary} resulting in a single pair with state $\ket{\Phi^+}_{13}$.
An example of this entanglement swapping protocol \cite{BriDur98:QSwapAndPur,Dur07:PuriAndApplications,Van08:Qrepeater} with $9$ nodes (i.e. $8$ hops) is depicted in Fig.~\ref{fig:ExampleEntanglementSwapping}. 

In general, this protocol deteriorates the quality of the entangled pair. 
To be precise, taking into account the same probability distributions for the pair used to teleport and for the pair in which one qubit has been teleported, we obtain a probability evolution of the characteristic parameters described by
\begin{equation}
\label{eq:SwappingProbEvol}
\begin{aligned}
A_{i+1} &= A_{i}^2 + B_i^2 + C_{i}^2 + D_i^2\\
B_{i+1} &= 2 A_{i} B_i + 2 C_{i} D_i\\
C_{i+1} &= 2 A_{i} C_i + 2 B_{i} D_i\\
D_{i+1} &= 2 A_{i} D_i + 2 C_{i} B_i\\
\end{aligned}
\end{equation}
where again $i$ represents the current step since the initial probability distribution in \eqref{eq:GenMixedState} evolves according to \eqref{eq:DeautschProbEvol} or \eqref{eq:SwappingProbEvol}. It is important to note that, after $\ell$ swapping steps we are able to cover $2^{\ell}$ link hops.

According to \eqref{eq:SwappingProbEvol}, the protocol tends to the stable and equiprobable configuration $A_i = B_i = C_i = D_i$. 
Hence, the distillation protocol has to compensate also for this fidelity loss in order to guarantee the target fidelity. 
For this reason, we propose a protocol name Burst $b$, where $b$ is the number of distillation steps done. 
In this protocol, we first schedule $b$ single link distillations (e.g., Burst $1$ in Fig~\ref{fig:ProtocolMessages}a and Burst $3$ in Fig~\ref{fig:ProtocolMessages}b). 
Consequently, we perform the necessary entanglement swap over these distilled pairs.
This protocol has the advantage that it significantly reduces the control signaling in the network since it uses only single-link distillation. 
In other words, it improves the latency of the communication.
Moreover, it snowballs the fidelity improvement given by distillation before letting the swapping protocol degrade the fidelity. 
This has the advantage to improve also the reliability compared to the protocol stack proposed in \cite{Van08:Qrepeater} where distillation and swapping are alternated in order to keep the fidelity above the target.
We performed extensive numerical investigations for several initial quantum states and, based on the results, we conjecture that the proposed approach is the optimal scheduling.
A formal proof of optimality, which seems non-trivial, is left as a future work.

Similarly to single link quantum communication protocols, we can adopt \ac{QEC} to construct a hybrid scheme capable of reliably transmitting the quantum information. From \eqref{eq:SwappingProbEvol} we observe that for small error probability (i.e., $\rho_i \ll 1$) and non-trivial unbalance in the particular error probabilities, we have that $B_{i+1} \approx 2 B_{i}$, $C_{i+1} \approx 2 C_{i}$, and $D_{i+1} \approx 2 D_{i}$. Then, we can state that entanglement swapping preserves the asymmetries of the equivalent quantum channel. As a consequence, for practical consideration, the asymmetry $\mathcal{A}_{\mathrm{eq}, i}$ is given by the initial state and the number of distillation steps.
For this reason, the entanglement swapping protocol affects the \ac{QEC} code design only by deteriorating the error parameter $\rho_i$.

\section{Numerical Results}
\label{sec:NumRes}
In this section we report the performance comparison of quantum hybrid communication protocols, using the logical qubit error probability $\rho_\mathrm{L}$, as given in \eqref{eq:PeGen3}.
To evaluate the parameter of \eqref{eq:PeGen3}, we firstly use \eqref{eq:DeautschProbEvol} and \eqref{eq:SwappingProbEvol} to find the final distilled pair state $\{A, B, C, D\}$, and then we find $\mathcal{A}_\mathrm{eq} = 2D/(B+C)$ and $\rho = B + C + D$.
In this way, we do not require Monte Carlo simulation to evaluate the results.
We report these comparisons for both the single link and network scenarios, varying the adopted $[[n,k]]$ with $(e_\mathrm{g}, e_\mathrm{Z})$ error correcting code. 
Due to implementation reasons, we focus our results on short quantum codes such as the $[[3,1]]$ and $[[5,1]]$ repetition codes, the $5$-qubit code \cite{Laf:96}, the $[[11,1]]$ code able to correct up to $2$ generic errors \cite{Gra07:Codes}, and the $[[9,1]]$ and $[[13,1]]$ asymmetric codes \cite{ChiVal:20a}.
In the following, we consider only the case where the initial state is a Werner one.

\subsection{Ad-hoc Coding over Single Quantum Link}

\begin{figure}[t]
    \centering
    \includegraphics[width = 0.7\columnwidth]{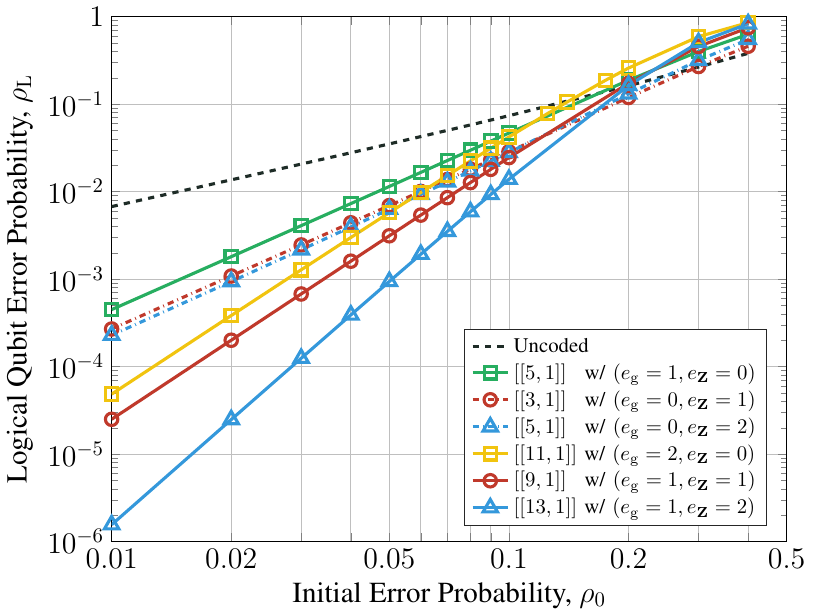}
    \caption{\justifying Logical qubit error probability against the initial error probability $\rho_0 = 1-\mathcal{F}_0$ considering a single distillation step. We consider bounded distance decoding for $[[n,k]]$ codes with $(e_\mathrm{g}, e_\mathrm{Z})$ error correction capability as reported in \eqref{eq:PeGen3}.}
    \label{fig:SingleLinkAsymCoding}
\end{figure}

In Fig.~\ref{fig:SingleLinkAsymCoding} we report the performance of a low-latency single quantum link communication varying the initial error probability $\rho_0$, where a single step of distillation is performed (see Fig.~\ref{fig:ProtocolMessages}a). 
The uncoded curve shows again that one distillation step does not significantly improve the fidelity when starting from a Werner state.
However, this apparently useless step is able to give us an asymmetry we can exploit by ad-hoc coding. 
In fact, adopting asymmetric \acp{QECC} (i.e., $e_\mathrm{Z} > 0$) we obtain an important performance boost. 
From the plot, we can observe some interesting behaviors.
Firstly, we note that the simple and easy-to-implement $[[3,1]]$ repetition code for a phase-flip channel can outperform the $5$-qubit code ($e_\mathrm{g} = 1$). 
In fact, in the presence of strong asymmetry, these two codes have the same effective error correction capability. 
Hence, the one with fewer qubits is able to outperform the other both in terms of error probability and codeword length. 
Secondly, considering the $[[5,1]]$ repetition code we observe no improvement compared to the $[[3,1]]$. This is due to the fact that the double $\M{Z}$ error is approaching the same occurrence probability of a single $\M{X}$, which is limiting further improvement. 
Similarly, the performance comparison between the $[[11,1]]$ and $[[9,1]]$ codes shows that it is inefficient to protect against generic errors when the asymmetry is sufficiently large.
Thirdly, we note that, for a fixed $e_\mathrm{g}$, we obtain an improvement by increasing $e_\mathrm{Z}$, as it is shown for the $[[9,1]]$ and $[[13,1]]$ asymmetric codes. 
This is not an obvious result, since to increase $e_\mathrm{Z}$ we have to increase the codeword length, making the overall performance depending also on the channel asymmetry. In the case of Fig.~\ref{fig:SingleLinkAsymCoding}, the channel asymmetry makes it advantageous to use codes with larger correction capability against $\M{Z}$ errors.  
Lastly, considering a target logical qubit error probability $\rho_\mathrm{L}^* = 10^{-3}$, while without codes we need an initial fidelity $\mathcal{F}_0 \ge 99.85 \%$, using the $[[3,1]]$ repetition code we require an initial fidelity $\mathcal{F}_0 \ge 98 \%$, and with the $[[13,1]]$ asymmetric code we require an initial fidelity $\mathcal{F}_0 \ge 95 \%$.

Another way to use Fig.~\ref{fig:SingleLinkAsymCoding}, is to perform code selection. Given the link parameter $\{ A, B, C, D\}$ and a number of available distilled pairs, using \eqref{eq:PeGen3} we can find the code with lowest $\rho_\mathrm{L}$ using at most the number of available pairs. Since the results are analytically derived, this does not require simulations. This analysis can be extended also to topological codes using minimum weight perfect matching decoders since analytical logical error rates was recently derived in~\cite{For24:XZZXRotAnalysis}.
Adopting other distillation algorithms, e.g. STRINGENT, it is possible to perform the same analysis numerically evaluating the corresponding $\rho_{i}$ and $\mathcal{A}_{\mathrm{eq}, i}$ of those protocols as a function of $\rho_0$. 
Then, using \eqref{eq:PeGen3} for each desired code, one can obtain a performance comparison similar to the one in Fig.~\ref{fig:SingleLinkAsymCoding}.
Whether an improvement is seen depends on the $\rho_i$ and $\mathcal{A}_{\mathrm{eq},i}$ of the specific distillation algorithm.

\subsection{Quantum Network Analysis}
\begin{figure}[t]
    \centering
    \includegraphics[width = 0.7\columnwidth]{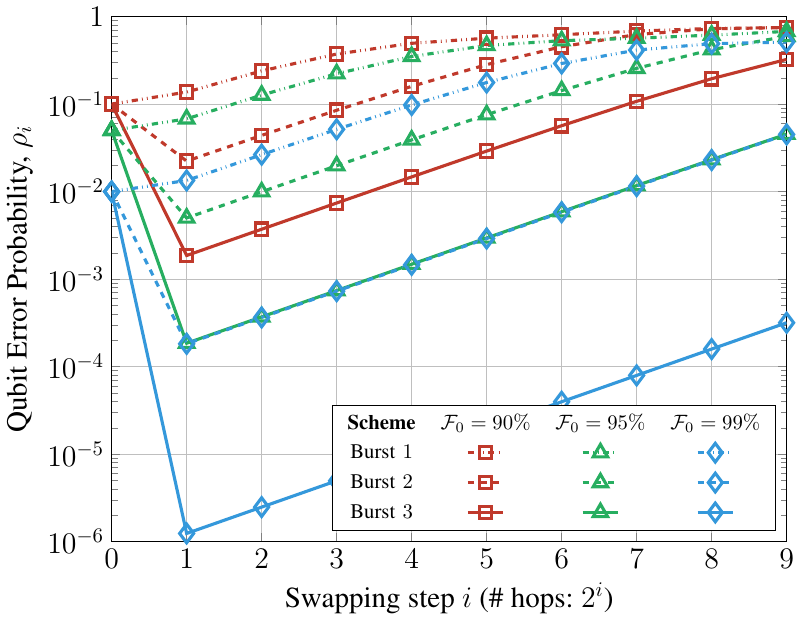}
    \caption{\justifying Evolution of the equivalent quantum teleportation channel error probability $\rho_i$ due to entanglement swapping \eqref{eq:SwappingProbEvol}, for Burst $b = 1,2,3$ protocols and initial fidelity $\mathcal{F}_0 = 90\%, 95\%, 99\%$.}
    \label{fig:SwappingWithBurst_Err}
\end{figure}

In Fig.~\ref{fig:SwappingWithBurst_Err} we report the effect on the qubit error probability $\rho_i$ in a quantum network scenario, varying the number of swapping steps $i$ and considering the Burst $b$ protocols proposed in Section~\ref{subsec:MultiLinks}.
As expected, when $\rho_1$ is sufficiently small we have that $\rho_i \approx 2^{i-1} \rho_1$. Then, to design an uncoded scheme it is sufficient to fix an $i_\mathrm{max}$ of steps we need to cover and act on $b$ in order to satisfy $\rho_\mathrm{L}^* \ge 2^{i_\mathrm{max}-1} \rho_1$. Note that the number of swapping steps is not the number of link hops. To be precise, we can cover $2^i$ hops with $i$ swapping step (see Fig.~\ref{fig:ExampleEntanglementSwapping}).
From the plot, we point out that Burst $1$ schemes, and in general all network schemes with just one distillation step, always give $\rho_i > \rho_0$, $i\geq 1$. 
This is because one distillation step gives a $2/3$ improvement factor, while the swapping gives a factor $2$ degradation, resulting, for non-trivial Werner states with $\rho_0 < 0.75$, in $\rho_1 \approx 4 \rho_0 / 3$ for low $\rho_0$.
Then, to gain an improvement in the qubit error probability, we have to adopt Burst $b$ schemes with $b>1$.
For example, considering an initial fidelity $\mathcal{F}_0 = 90 \%$ we have $\rho_i < \rho_0$ until swapping step $3$ and step $6$ for Burst $2$ and Burst $3$ protocols.

\begin{figure}[t]
    \centering
    \includegraphics[width = 0.7\columnwidth]{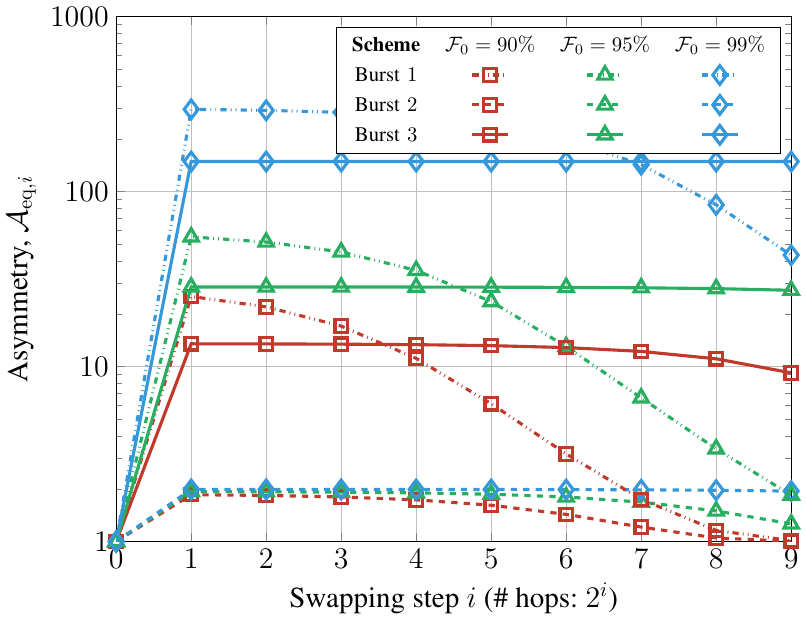}
    \caption{\justifying Evolution of the quantum teleportation channel equivalent asymmetry $\mathcal{A}_{\mathrm{eq},i}$ due to entanglement swapping \eqref{eq:SwappingProbEvol}, for Burst $b = 1,2,3$ protocols and initial fidelity $\mathcal{F}_0 = 90\%, 95\%, 99\%$. The asymmetry $\mathcal{A}_\mathrm{eq} = 1$ represents the well-known depolarizing channel.}
    \label{fig:SwappingWithBurst_Asym}
\end{figure}

In Fig.~\ref{fig:SwappingWithBurst_Asym} we report the effect on the equivalent asymmetry $\mathcal{A}_{\mathrm{eq},i}$ in a quantum network scenario, varying the number of swapping steps $i$ for Burst $b$ protocols.
As expected from Fig.~\ref{fig:WernerStateEvolution}, we have that Burst~$1$ and Burst~$3$ give good asymmetry values for ad-hoc coding, while Burst $2$ is almost symmetric ($p_\mathrm{Z} = p_\mathrm{X}$ as can be observed from Fig.~\ref{fig:WernerStateEvolution}).
For this reason, even if Burst~$2$ can already provide $\rho_i < \rho_0$ for some values of $i$, it can be useful to use Burst~$3$ aiming to adopt ad-hoc \ac{QEC} over the distilled pairs.
Moreover, we remind that in terms of latency, Burst~$2$ and Burst~$3$ are equivalent as shown in Fig.~\ref{fig:ProtocolMessages}b.
As previously discussed in Section~\ref{subsec:MultiLinks}, for $\rho_i$ of practical interest (i.e., $\rho_i < 0.1$), we can see that the asymmetry is preserved when the swapping protocol is performed.

\begin{figure}[t]
    \centering
    \includegraphics[width = 0.7\columnwidth]{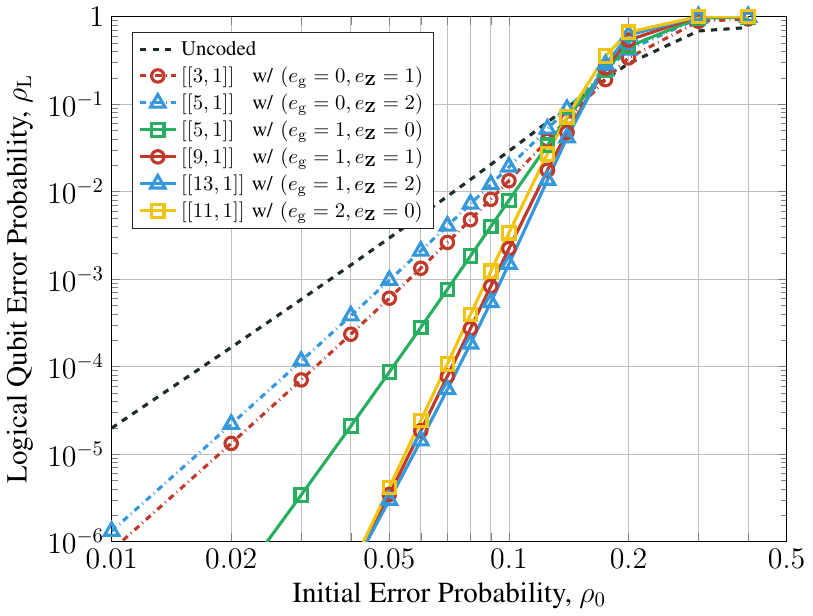}
    \caption{\justifying Logical qubit error probability against the initial error probability $\rho_0 = 1-\mathcal{F}_0$ considering the Burst $3$ protocol and $5$ swapping steps (i.e., $32$ network hops). We consider bounded distance decoding for $[[n,k]]$ codes with $(e_\mathrm{g}, e_\mathrm{Z})$ error correction capability as reported in \eqref{eq:PeGen3}.}
    \label{fig:NetworkAsymCoding}
\end{figure}
\subsection{Ad-hoc Coding over Quantum Network}
In Fig.~\ref{fig:NetworkAsymCoding} we report the performance of communication over a quantum network by varying the initial error probability $\rho_0$, where Burst $3$ is adopted and $5$ swapping steps are performed.
Regarding the comparison with Fig.~\ref{fig:SingleLinkAsymCoding}, in this case, \acp{QECC} act on a different regime. 
In fact, comparing channel parameters for $1$ and $3$ steps of distillation we have that the asymmetries $\mathcal{A}_{\mathrm{eq},i}$ are similar (see also Fig.~\ref{fig:SwappingWithBurst_Asym}), but the error probabilities $\rho_i$ are much smaller when accounting for $3$ distillation steps.
For this reason, the symmetric $5$-qubit code outperforms the repetition codes designed for phase-flip errors. Larger codes with $n \geq 9$ give further improvements, as reported in the figure. More precisely, since the channel is asymmetric, we can suitably design asymmetric codes, balancing $e_\mathrm{g}$ and $e_\mathrm{Z}$. 
This can be easily seen from the plot, observing that the $[[11, 1]]$ symmetric code performs practically the same as the $[[9,1]]$ asymmetric code. The latter code, requiring fewer qubits per codeword, could be thus preferred for practical implementations.  
In general, we emphasize that an estimation of the initial parameters can be exploited for a tailored \ac{QECC} design. These parameters could be taken from the nominal one given by the adopted technology. However, when these parameters are not available, asymmetry cannot be exploited and the only possibility to reduce the protocol latency, compared to distillation-only schemes, is to construct sub-optimal hybrid schemes using symmetric \acp{QECC}. 

\section{Discussion and Future Works}\label{sec:discussion}
In this work, our focus have lied on low-latency protocols, demonstrating that good performance can also be achieved when few steps of distillation are considered. In the current vision, first generation networks will adopt only distillation to make the communication reliable, while third generation networks will adopt only quantum error correction to have both reliable and low latency communication protocols \cite{Mur16:QnetworkGenerations, irtf-qirg-principles-11}. 
In our envisioned trajectory, positioned between these two generations, hybrid schemes could emerge, leveraging distillation while also addressing latency concerns.

One possible direction for future works could deepen the discussion on the advantages of combining distillation and \ac{QEC} instead of directly distilling perfect \ac{EPR} pairs.
Such an exploration could provide valuable insights into the conditions under which quantum networks can attain sufficient reliability, thereby enabling a shift from addressing solely reliability through distillation to addressing broader concerns (e.g., latency), marking the transition from first-generation to second-generation networks.
This can be done by conducting a comparative analysis across schemes employing varying numbers of distillation steps. 
In addition to the provided framework, to have a comprehensive analysis for protocols with different latency, it is necessary to take into account that the scheme with the larger latency will: 
$i)$ suffer more decoherence; 
$ii)$ keep the communication link busy for more time;
$iii)$ slow down the applications requiring the teleportation.

Another possible direction is the assessment of the asymmetry, a crucial aspect in order to choose a proper coding strategy. 
Quantum communication links could be: $i)$ non-adaptive; $ii)$ adaptive.
In non-adaptive quantum communication links, we can choose a proper coding scheme based on the adopted technology for distributing entanglement. In fact, given the technology, we have the initial state probability distribution $\{A_0, B_0, C_0, D_0\}$. Having a fidelity target (or equivalently a reliability target), we can therefore execute our analysis to find the final asymmetry and the best solution.
In adaptive quantum communication links, we could imagine a protocol which periodically estimates $\{A_0, B_0, C_0, D_0\}$ (e.g., via tomography), and then adapts the coding scheme, accordingly.
\section{Conclusions}\label{sec:conclusions}
We proposed a hybrid schemes where distillation is used together with asymmetric \acp{QECC} to exchange information over a quantum network. The key idea is that distillation gives rise to an equivalent quantum communication channel with asymmetric Pauli error probabilities. We show that significant performance improvements can be achieved by performing a few steps of distillations, followed by teleporting protected with \acp{QECC}. 
For example, starting from a Werner state, assuming one distillation step and a target logical qubit error probability $10^{-3}$, the distillation-only protocol requires an initial fidelity $\mathcal{F}_0 \ge 99.85 \%$, while the proposed hybrid protocol using an asymmetric $[[13,1]]$ code works with $\mathcal{F}_0 \ge 95 \%$.
In a quantum network scenario where several swapping steps are needed to connect two nodes, we have shown how to design efficient hybrid schemes fulfilling given network quality requirements. 
It results that, for both single link and quantum network scenarios, the proposed hybrid distillation/\ac{QEC} protocols give substantial advantages in terms of latency and fidelity with respect to distillation-only communication protocols. 




\bibliographystyle{IEEEtran}
\bibliography{Files/IEEEabrv,Files/StringDefinitions,Files/StringDefinitions2,Files/refs}

\begin{thebibliography}{10}
\providecommand{\url}[1]{#1}
\csname url@samestyle\endcsname
\providecommand{\newblock}{\relax}
\providecommand{\bibinfo}[2]{#2}
\providecommand{\BIBentrySTDinterwordspacing}{\spaceskip=0pt\relax}
\providecommand{\BIBentryALTinterwordstretchfactor}{4}
\providecommand{\BIBentryALTinterwordspacing}{\spaceskip=\fontdimen2\font plus
\BIBentryALTinterwordstretchfactor\fontdimen3\font minus
  \fontdimen4\font\relax}
\providecommand{\BIBforeignlanguage}[2]{{%
\expandafter\ifx\csname l@#1\endcsname\relax
\typeout{** WARNING: IEEEtran.bst: No hyphenation pattern has been}%
\typeout{** loaded for the language `#1'. Using the pattern for}%
\typeout{** the default language instead.}%
\else
\language=\csname l@#1\endcsname
\fi
#2}}
\providecommand{\BIBdecl}{\relax}
\BIBdecl

\bibitem{WehElkHan:18}
S.~Wehner, D.~Elkouss, and R.~Hanson, ``Quantum internet: A vision for the road
  ahead,'' \emph{Science}, vol. 362, no. 6412, 2018.

\bibitem{Cac19:QInternet}
A.~S. Cacciapuoti, M.~Caleffi, F.~Tafuri, F.~S. Cataliotti, S.~Gherardini, and
  G.~Bianchi, ``Quantum internet: networking challenges in distributed quantum
  computing,'' \emph{IEEE Network}, vol.~34, no.~1, pp. 137--143, 2019.

\bibitem{Pom22:experimentalQI}
M.~Pompili, C.~Delle~Donne, I.~te~Raa, B.~van~der Vecht, M.~Skrzypczyk,
  G.~Ferreira, L.~de~Kluijver, A.~J. Stolk, S.~L. Hermans, P.~Pawe{\l}czak
  \emph{et~al.}, ``Experimental demonstration of entanglement delivery using a
  quantum network stack,'' \emph{npj Quantum Information}, vol.~8, no.~1, p.
  121, 2022.

\bibitem{BenBra84:QKD}
C.~H. Bennett and G.~Brassard, ``Quantum cryptography: Public key distribution
  and coin tossing,'' \emph{arXiv preprint arXiv:2003.06557}, 2020.

\bibitem{BenWie92:SuperDenseCoding}
C.~H. Bennett and S.~J. Wiesner, ``Communication via one- and two-particle
  operators on {E}instein-{P}odolsky-{R}osen states,'' \emph{Physical review
  letters}, vol.~69, no.~20, p. 2881, 1992.

\bibitem{Bal22:MultiCoreQComp}
H.~Jnane, B.~Undseth, Z.~Cai, S.~C. Benjamin, and B.~Koczor, ``Multicore
  quantum computing,'' \emph{Physical Review Applied}, vol.~18, no.~4, 2022.

\bibitem{SunGup22:QcircuitsOverQNet}
R.~G. Sundaram, H.~Gupta, and C.~Ramakrishnan, ``Distribution of quantum
  circuits over general quantum networks,'' in \emph{2022 IEEE International
  Conference on Quantum Computing and Engineering (QCE)}.\hskip 1em plus 0.5em
  minus 0.4em\relax IEEE, 2022, pp. 415--425.

\bibitem{Ferrari2023:DistrQC}
D.~Ferrari, S.~Carretta, and M.~Amoretti, ``A modular quantum compilation
  framework for distributed quantum computing,'' \emph{IEEE Transactions on
  Quantum Engineering}, vol.~4, pp. 1--13, 2023.

\bibitem{Deg17:ReviewQSensing}
C.~L. Degen, F.~Reinhard, and P.~Cappellaro, ``Quantum sensing,'' \emph{Reviews
  of modern physics}, vol.~89, no.~3, p. 035002, 2017.

\bibitem{Bi19:QRemoteSensing}
S.~Bi, ``Research on quantum remote sensing science and technology,'' in
  \emph{Infrared Remote Sensing and Instrumentation XXVII}.\hskip 1em plus
  0.5em minus 0.4em\relax SPIE, 2019.

\bibitem{irtf-qirg-principles-11}
\BIBentryALTinterwordspacing
W.~Kozlowski, S.~Wehner, R.~V. Meter, and et. al., ``{Architectural Principles
  for a Quantum Internet},'' Internet Engineering Task Force, Internet-Draft
  draft-irtf-qirg-principles-11, Aug. 2022, work in Progress. [Online].
  Available:
  \url{https://datatracker.ietf.org/doc/draft-irtf-qirg-principles/11/}
\BIBentrySTDinterwordspacing

\bibitem{Ben93:teleporting}
C.~H. Bennett, G.~Brassard, C.~Cr{\'e}peau, R.~Jozsa, A.~Peres, and W.~K.
  Wootters, ``Teleporting an unknown quantum state via dual classical and
  {E}instein-{P}odolsky-{R}osen channels,'' \emph{Physical review letters},
  vol.~70, no.~13, 1993.

\bibitem{Fow10:QcommWithSurf}
A.~G. Fowler, D.~S. Wang, C.~D. Hill, T.~D. Ladd, R.~Van~Meter, and L.~C.
  Hollenberg, ``Surface code quantum communication,'' \emph{Physical review
  letters}, vol. 104, no.~18, p. 180503, 2010.

\bibitem{GreHorZei89:GHZ}
D.~M. Greenberger, M.~A. Horne, and A.~Zeilinger, ``Going beyond bell’s
  theorem,'' \emph{Bell’s theorem, quantum theory and conceptions of the
  universe}, pp. 69--72, 1989.

\bibitem{Hon04:GHZAndWStates}
E.~D'Hondt and P.~Panangaden, ``The computational power of the {W} and {GHZ}
  states,'' \emph{arXiv preprint quant-ph/0412177}, 2004.

\bibitem{Pom21:ExperimentalQNetwork}
M.~Pompili, S.~L.~N. Hermans, S.~Baier, H.~K.~C. Beukers, P.~C. Humphreys,
  R.~N. Schouten, R.~F.~L. Vermeulen, M.~J. Tiggelman, L.~dos Santos~Martins,
  B.~Dirkse, S.~Wehner, and R.~Hanson, ``Realization of a multinode quantum
  network of remote solid-state qubits,'' \emph{Science}, vol. 372, no. 6539,
  pp. 259--264, apr 2021.

\bibitem{Cir97:EntanGen}
J.~I. Cirac, P.~Zoller, H.~J. Kimble, and H.~Mabuchi, ``Quantum state transfer
  and entanglement distribution among distant nodes in a quantum network,''
  \emph{Physical Review Letters}, vol.~78, no.~16, p. 3221, 1997.

\bibitem{Chi05:EntanGen}
L.~Childress, J.~Taylor, A.~S. S{\o}rensen, and M.~D. Lukin, ``Fault-tolerant
  quantum repeaters with minimal physical resources and implementations based
  on single-photon emitters,'' \emph{Physical Review A}, vol.~72, no.~5, 2005.

\bibitem{Loo06:EntaGen}
P.~Van~Loock, T.~Ladd, K.~Sanaka, F.~Yamaguchi, K.~Nemoto, W.~Munro, and
  Y.~Yamamoto, ``Hybrid quantum repeater using bright coherent light,''
  \emph{Physical review letters}, vol.~96, no.~24, p. 240501, 2006.

\bibitem{Uph16:EntaGen}
M.~Uphoff, M.~Brekenfeld, G.~Rempe, and S.~Ritter, ``An integrated quantum
  repeater at telecom wavelength with single atoms in optical fiber cavities,''
  \emph{Applied Physics B}, vol. 122, pp. 1--15, 2016.

\bibitem{Hu21:EntaGen}
X.-M. Hu, C.-X. Huang, Y.-B. Sheng, L.~Zhou, B.-H. Liu, Y.~Guo, C.~Zhang, W.-B.
  Xing, Y.-F. Huang, C.-F. Li \emph{et~al.}, ``Long-distance entanglement
  purification for quantum communication,'' \emph{Physical review letters},
  vol. 126, no.~1, p. 010503, 2021.

\bibitem{RoadmapIntegratedQPhotonics22}
G.~Moody, V.~J. Sorger, D.~J. Blumenthal, P.~W. Juodawlkis, W.~Loh,
  C.~Sorace-Agaskar, A.~E. Jones, K.~C. Balram, J.~C. Matthews, A.~Laing
  \emph{et~al.}, ``2022 roadmap on integrated quantum photonics,''
  \emph{Journal of Physics: Photonics}, vol.~4, no.~1, p. 012501, 2022.

\bibitem{Ben96:purification}
C.~H. Bennett, G.~Brassard, S.~Popescu, B.~Schumacher, J.~A. Smolin, and W.~K.
  Wootters, ``Purification of noisy entanglement and faithful teleportation via
  noisy channels,'' \emph{Physical review letters}, vol.~76, no.~5, p. 722,
  1996.

\bibitem{Deu96:purification}
D.~Deutsch, A.~Ekert, R.~Jozsa, C.~Macchiavello, S.~Popescu, and A.~Sanpera,
  ``Quantum privacy amplification and the security of quantum cryptography over
  noisy channels,'' \emph{Physical review letters}, vol.~77, no.~13, p. 2818,
  1996.

\bibitem{CacCalVan:20}
A.~S. Cacciapuoti, M.~Caleffi, R.~Van~Meter, and L.~Hanzo, ``When entanglement
  meets classical communications: Quantum teleportation for the quantum
  {Internet},'' \emph{{IEEE} Trans. Commun.}, vol.~68, no.~6, 2020.

\bibitem{Van08:Qrepeater}
R.~Van~Meter, T.~D. Ladd, W.~J. Munro, and K.~Nemoto, ``System design for a
  long-line quantum repeater,'' \emph{IEEE/ACM Transactions On Networking},
  vol.~17, no.~3, pp. 1002--1013, 2008.

\bibitem{Weh19:QStackCoopHeralding}
A.~Dahlberg, M.~Skrzypczyk, T.~Coopmans, L.~Wubben, F.~Rozpedek, M.~Pompili,
  A.~Stolk, P.~Pawe{\l}czak, R.~Knegjens, J.~de~Oliveira~Filho \emph{et~al.},
  ``A link layer protocol for quantum networks,'' in \emph{Proceedings of the
  ACM special interest group on data communication}, 2019, pp. 159--173.

\bibitem{KozWeh20:QStackCoopHeralding}
W.~Kozlowski, A.~Dahlberg, and S.~Wehner, ``Designing a quantum network
  protocol,'' in \emph{Proceedings of the 16th international conference on
  emerging networking experiments and technologies}, 2020, pp. 1--16.

\bibitem{Sho94:Factoring}
P.~W. {Shor}, ``Algorithms for quantum computation: discrete logarithms and
  factoring,'' in \emph{Proceedings 35th Annual Symposium on Foundations of
  Computer Science}, Nov 1994, pp. 124--134.

\bibitem{BraKit98:Surface}
S.~B. Bravyi and A.~Y. Kitaev, ``Quantum codes on a lattice with boundary,''
  \emph{arXiv preprint quant-ph/9811052}, 1998.

\bibitem{ValFor23:Surface}
L.~Valentini, D.~Forlivesi, and M.~Chiani, ``Performance analysis of quantum
  error-correcting surface codes over asymmetric channels,'' in \emph{Proc.
  2023 IEEE Int. Conf. Commun.}, Rome, Italy, May 2023.

\bibitem{For24:XZZXRotAnalysis}
D.~Forlivesi, L.~Valentini, and M.~Chiani, ``Logical error rates of {XZZX} and
  rotated quantum surface codes,'' \emph{{IEEE} J. Sel. Areas Commun.}, 2024.

\bibitem{ChiConWin:20}
M.~Chiani, A.~Conti, and M.~Z. Win, ``Piggybacking on quantum streams,''
  \emph{Physical Review A}, vol. 102, no.~1, jul 2020.

\bibitem{Sho:95}
P.~W. Shor, ``Scheme for reducing decoherence in quantum computer memory,''
  \emph{Phys. Rev. A}, vol.~52, pp. R2493--R2496, Oct 1995.

\bibitem{Laf:96}
R.~Laflamme, C.~Miquel, J.~P. Paz, and W.~H. Zurek, ``Perfect quantum error
  correcting code,'' \emph{Physical Review Letters}, vol.~77, no.~1, 1996.

\bibitem{Gra07:Codes}
M.~Grassl, ``{Bounds on the minimum distance of linear codes and quantum
  codes},'' Online available at \url{http://www.codetables.de}, 2007, accessed
  on 2019-12-20.

\bibitem{Sar:2009}
P.~K. Sarvepalli, A.~Klappenecker, and M.~R{\"o}tteler, ``Asymmetric quantum
  codes: constructions, bounds and performance,'' \emph{Proceedings of the
  Royal Society A: Mathematical, Physical and Engineering Sciences}, vol. 465,
  no. 2105, pp. 1645--1672, 2009.

\bibitem{ChiVal:20a}
M.~Chiani and L.~Valentini, ``Short codes for quantum channels with one
  prevalent {Pauli} error type,'' \emph{{IEEE} J. Sel. Areas Inf. Theory},
  vol.~1, no.~2, pp. 480--486, 2020.

\bibitem{Ste96:CSS}
A.~Steane, ``Multiple-particle interference and quantum error correction,''
  \emph{Proceedings of the Royal Society of London. Series A}, vol. 452, no.
  1954, pp. 2551--2577, 1996.

\bibitem{CalSho96:CSS}
A.~R. Calderbank and P.~W. Shor, ``Good quantum error-correcting codes exist,''
  \emph{Physical Review A}, vol.~54, no.~2, p. 1098, 1996.

\bibitem{Ste99:CSS}
A.~M. Steane, ``Enlargement of calderbank-shor-steane quantum codes,''
  \emph{{IEEE} Trans. Inf. Theory}, vol.~45, no.~7, pp. 2492--2495, 1999.

\bibitem{Mur16:QnetworkGenerations}
S.~Muralidharan, L.~Li, J.~Kim, N.~L{\"u}tkenhaus, M.~D. Lukin, and L.~Jiang,
  ``Optimal architectures for long distance quantum communication,''
  \emph{Scientific reports}, vol.~6, no.~1, p. 20463, 2016.

\bibitem{Hei04:MultipartyGraphStates}
M.~Hein, J.~Eisert, and H.~J. Briegel, ``Multiparty entanglement in graph
  states,'' \emph{Physical Review A}, vol.~69, no.~6, p. 062311, 2004.

\bibitem{Hei06:GraphStates}
M.~Hein, W.~D{\"u}r, J.~Eisert, R.~Raussendorf, M.~Nest, and H.-J. Briegel,
  ``Entanglement in graph states and its applications,'' \emph{arXiv preprint
  quant-ph/0602096}, 2006.

\bibitem{Mei19:DistribGraphStates}
C.~Meignant, D.~Markham, and F.~Grosshans, ``Distributing graph states over
  arbitrary quantum networks,'' \emph{Physical Review A}, vol. 100, no.~5, p.
  052333, 2019.

\bibitem{Rie00:introduction}
E.~Rieffel and W.~Polak, ``An introduction to quantum computing for
  non-physicists,'' \emph{ACM Computing Surveys (CSUR)}, vol.~32, no.~3, pp.
  300--335, 2000.

\bibitem{NieChu:10}
M.~A. Nielsen and I.~L. Chuang, \emph{Quantum Computation and Quantum
  Information}.\hskip 1em plus 0.5em minus 0.4em\relax Cambridge University
  Press, 2010.

\bibitem{Wer89:WernerState}
R.~F. Werner, ``Quantum states with {E}instein-{P}odolsky-{R}osen correlations
  admitting a hidden-variable model,'' \emph{Physical Review A}, vol.~40,
  no.~8, p. 4277, 1989.

\bibitem{Zha02:WernerPrep}
Y.-S. Zhang, Y.-F. Huang, C.-F. Li, and G.-C. Guo, ``Experimental preparation
  of the werner state via spontaneous parametric down-conversion,''
  \emph{Physical Review A}, vol.~66, no.~6, p. 062315, 2002.

\bibitem{Das21:IBMQPuri}
S.~Das, M.~S. Rahman, and M.~Majumdar, ``Design of a quantum repeater using
  quantum circuits and benchmarking its performance on an ibm quantum
  computer,'' \emph{Quantum Information Processing}, vol.~20, 2021.

\bibitem{Dur99:NoisyPurif}
W.~D{\"u}r, H.-J. Briegel, J.~I. Cirac, and P.~Zoller, ``Quantum repeaters
  based on entanglement purification,'' \emph{Physical Review A}, vol.~59,
  no.~1, 1999.

\bibitem{Dur07:PuriAndApplications}
W.~D{\"u}r and H.~J. Briegel, ``Entanglement purification and quantum error
  correction,'' \emph{Reports on Progress in Physics}, vol.~70, no.~8, p. 1381,
  2007.

\bibitem{ChiSim22:LearningPurifNoisyClasMsg}
H.~H.~S. Chittoor and O.~Simeone, ``Learning quantum entanglement distillation
  with noisy classical communications,'' \emph{arXiv preprint
  arXiv:2205.08561}, 2022.

\bibitem{Mac98:ProofPurification}
C.~Macchiavello, ``On the analytical convergence of the {QPA} procedure,''
  \emph{Physics Letters A}, vol. 246, no.~5, pp. 385--388, 1998.

\bibitem{KraAlbJia2019:OptPurification}
S.~Krastanov, V.~V. Albert, and L.~Jiang, ``Optimized entanglement
  purification,'' \emph{Quantum}, vol.~3, p. 123, 2019.

\bibitem{Goo2023:nkPurification}
K.~Goodenough, S.~de~Bone, V.~L. Addala, S.~Krastanov, S.~Jansen, D.~Gijswijt,
  and D.~Elkouss, ``Near-term $n$ to $k$ distillation protocols using graph
  codes,'' \emph{arXiv preprint arXiv:2303.11465}, 2023.

\bibitem{Got:09}
D.~Gottesman, ``An introduction to quantum error correction and fault-tolerant
  quantum computation,'' \emph{arXiv preprint quant-ph/0904.2557}, 2009.

\bibitem{Mac04:QLDPC}
D.~J. MacKay, G.~Mitchison, and P.~L. McFadden, ``Sparse-graph codes for
  quantum error correction,'' \emph{IEEE Transactions on Information Theory},
  vol.~50, no.~10, pp. 2315--2330, 2004.

\bibitem{Nickerson2013}
N.~H. Nickerson, Y.~Li, and S.~C. Benjamin, ``Topological quantum computing
  with a very noisy network and local error rates approaching one percent,''
  \emph{Nature Communications}, vol.~4, no.~1, p. 1756, Apr 2013.

\bibitem{Cac20:QTPforQI}
A.~S. Cacciapuoti, M.~Caleffi, R.~Van~Meter, and L.~Hanzo, ``When entanglement
  meets classical communications: Quantum teleportation for the quantum
  internet,'' \emph{{IEEE} Trans. Commun.}, vol.~68, no.~6, pp. 3808--3833,
  2020.

\bibitem{Bar05:DetectorEntanGen}
S.~D. Barrett and P.~Kok, ``Efficient high-fidelity quantum computation using
  matter qubits and linear optics,'' \emph{Physical Review A}, vol.~71, no.~6,
  p. 060310, 2005.

\bibitem{SteAnd:96}
A.~M. Steane, ``Error correcting codes in quantum theory,'' \emph{Physical
  Review Letters}, vol.~77, no.~5, p. 793, 1996.

\bibitem{BriDur98:QSwapAndPur}
H.-J. Briegel, W.~D{\"u}r, J.~I. Cirac, and P.~Zoller, ``Quantum repeaters: the
  role of imperfect local operations in quantum communication,'' \emph{Physical
  Review Letters}, vol.~81, no.~26, 1998.

\end{thebibliography}


\end{document}